\RequirePackage[mathlines]{lineno} 
\documentclass[a4paper,11pt]{article}

\usepackage{jheppub} 

\usepackage[T1]{fontenc} 

\usepackage{threeparttable}
\usepackage{multirow}
\usepackage{multicol}
\usepackage{amssymb,bm,mathrsfs,bbm,amscd}
\usepackage{float}
\usepackage{subcaption}
\usepackage{overpic}
\usepackage{feynmp}
\usepackage{tikz-feynman}
\tikzfeynmanset{compat=1.1.0}

\let\oldequation\equation
\let\oldendequation\endequation
\renewenvironment{equation}
  {\linenomathNonumbers\oldequation}
  {\oldendequation\endlinenomath}

\def \ee   {e^{+}e^{-}}
\def \piz  {\pi^{0}}
\def \pip  {\pi^{+}}
\def \pim  {\pi^{-}}
\def \gam  {\gamma}
\def \sgmp {\Sigma^{+}}
\def \Ks {K_{S}^{0}}

\def \ifb  {\mbox{fb$^{-1}$}}
\def \gev  {\mbox{GeV}}
\def \gevc {\mbox{GeV/$c$}}
\def \gevcc {\mbox{GeV/$c^2$}}
\def \mev  {\mbox{MeV}}

\def \csq  {c^{2}}
\def \cfsq {c^{4}}

\def \romanOne   {\uppercase\expandafter{\romannumeral1}}
\def \romanTwo   {\uppercase\expandafter{\romannumeral2}}
\def \romanThree {\uppercase\expandafter{\romannumeral3}}
\def \romanFour  {\uppercase\expandafter{\romannumeral4}}
\def \romanFive  {\uppercase\expandafter{\romannumeral5}}
\def \romanSix   {\uppercase\expandafter{\romannumeral6}}
\def \romanSeven {\uppercase\expandafter{\romannumeral7}}
\def \romanEight {\uppercase\expandafter{\romannumeral8}}

\def \mbc {M_{\rm{BC}}}
\def \dE {\Delta E}
\def \ebeam {E_{\mathrm{beam}}}

\def \lcp {\Lambda_{c}^{+}}
\def \lcm {\bar{\Lambda}_{c}^{-}}
\def \lcplcm {\Lambda_{c}^{+}\bar{\Lambda}_{c}^{-}}

\begin{document}

\title{\boldmath Measurement of the branching fractions of the singly Cabibbo-suppressed decays $\lcp\to p\eta$ and $\lcp\to p\omega$}

\collaborationImg{\includegraphics[width=0.15\textwidth, angle=90]{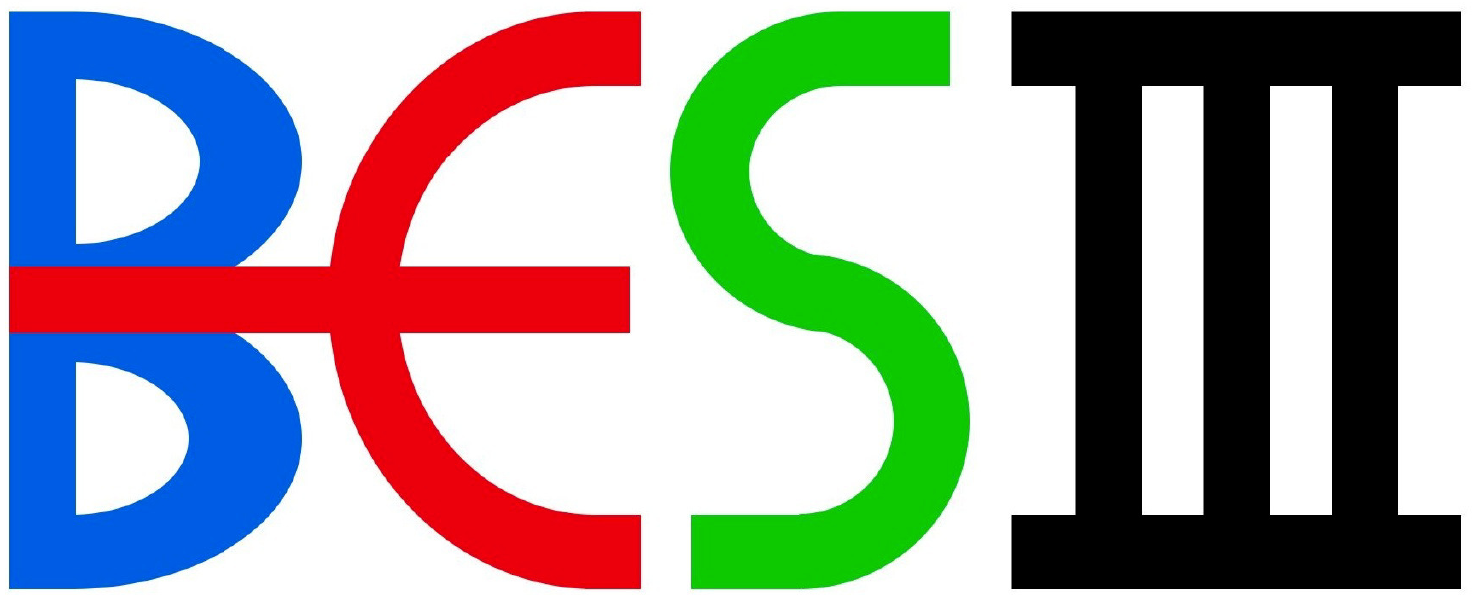}}
\collaboration{The BESIII collaboration}
\emailAdd{besiii-publications@ihep.ac.cn}

\abstract{
Based on 4.5 $\mbox{fb$^{-1}$}$ $\ee$ collision data collected with BESIII detector at seven energy points between 4.600 and 4.699 GeV, the branching fractions for $\lcp\to p\eta$ and $\lcp\to p\omega$ were measured by means of single-tag method. The branching fractions of $\lcp\to p\eta$ and $\lcp\to p\omega$ are determined to be $(1.57\pm0.11_{\rm {stat}}\pm0.04_{\rm{syst}})\times10^{-3}$ and $(1.11\pm0.20_{\rm{stat}}\pm0.07_{\rm{syst}})\times10^{-3}$, with a statistical significance of greater than 10 $\sigma$ and 5.7~$\sigma$, respectively. These results are consistent with the previous measurements by BESIII, LHCb and Belle, and the result of $\lcp\to p\eta$ is the most precise to date.
}


\arxivnumber{00000.00000}

\maketitle
\flushbottom

\section{Introduction}
\label{sec:introduction}
\hspace{1.5em} 
Studies on weak decays of charmed baryons provide crucial information on the strong and weak interactions in the charm sector.  Decay amplitudes of charmed hadrons can be divided into factorizable terms, where the strong and weak parts can be treated separately,
and non-factorizable terms, where the two parts are entangled and hard to calculate~\cite{Chau:1982da, Chau:1986jb, Chau:1995gk}. 
In the case of charmed mesons, non-factorizable terms are negligible compared to factorizable ones. However, for charmed baryon decays, the non-factorizable terms, such as inner exchange and emission of $W$ bosons, are no longer negligible and increase the complexity of calculation.  For example, $W$-exchange manifesting a pole diagram is no longer subject to helicity and color suppression~\cite{Cheng:1993gf}.  This makes theoretical calculations of the charmed baryon decays more challenging than those of charmed mesons.  There are many theoretical models and approaches dealing with charmed baryon decays, including the covariant confined quark model~\cite{Korner:1978ec,Korner:1992wi,Ivanov:1997ra}, pole model~\cite{Cheng:1991sn,Xu:1992vc,Cheng:1993gf,Xu:1992sw,Zenczykowski:1993jm,Sharma:1998rd}, current algebra~\cite{Cheng:1993gf,Sharma:1998rd,Uppal:1994pt}, and SU(3) flavor symmetry~\cite{Sharma:1996sc}.  To discriminate between these different models, experimental measurements are necessary.

There has been much progress in measurements of decay rates~\cite{Cheng2021,Cheng:2021vca} of the charmed baryons. Using a data set at $\sqrt{s}=4.600~\rm{GeV}$, which is just above the threshold of $\lcplcm$ pair production, BESIII improved the precision of the branching fractions of the Cabibbo-favoured $\lcp$ decays~\cite{BESIII:2015bjk,BESIII:2016yrc,BESIII:2015ysy,Li:2021iwf}. However, the experimental precision of Cabibbo-suppressed $\lcp$ decay rates, such as for the decays of $\lcp\to p\eta$ and $p\omega$, remains poor. BESIII measured the branching fraction of $\lcp\to p\eta$ in 2017~\cite{BESIII:2017fim} with a statistical significance of less than 5$\sigma$. LHCb measured the relative branching fraction of $\lcp\rightarrow p\omega$ in the same year with greater than 5$\sigma$ significance~\cite{LHCb:2017yqf}, where $\omega$ was reconstructed through the leptonic channel $\omega\to\mu^{+}\mu^{-}$. Belle also reported the relative branching fractions of both $\lcp\to p\eta$ and $\lcp\to p\omega$ in 2021 with a higher precision~\cite{Belle:2021mvw, Belle:2021btl}. All these measurements are listed in \tablename~\ref{tab:experiTheoPetaPomega}.

\begin{table}[tp]\footnotesize
	\centering
	\caption{Measurements and predictions of the branching fractions of $\lcp\rightarrow p\eta$ and $\lcp\to p\omega$ (in units of $10^{-3}$) from different experiments and theoretical calculations. The superscript $a$($b$) denotes the assumption of $P$-wave amplitude of $\lcp\rightarrow \Xi^{0}K^{+}$ is positive(negative). The superscript $c$($d$) denotes SU(3) flavor symmetry is conserved(broken).}
	\label{tab:experiTheoPetaPomega}
	\begin{threeparttable}
		\begin{tabular}{c|c|c|c}
			\hline\hline
			  \multicolumn{2}{c|}{}  & $\mathcal{B}(\lcp\rightarrow p\eta$) & $\mathcal{B}(\lcp\to p\omega)$ \\
			\hline
			\multicolumn{2}{c|}{BESIII} & $1.24\pm0.28\pm0.10$~\cite{BESIII:2017fim} & -\\
			\hline
			\multicolumn{2}{c|}{LHCb} & - & $0.94\pm0.32\pm0.22$~\cite{LHCb:2017yqf} \\
			\hline
			\multicolumn{2}{c|}{Belle} & $1.42\pm0.05\pm0.11$~\cite{Belle:2021mvw} & $0.827\pm0.075\pm0.075$~\cite{Belle:2021btl}\\
			\hline
            \multicolumn{2}{c|}{This paper} & $1.57\pm0.11\pm0.04$ & $1.11\pm0.20\pm0.07$ \\
            \hline
			\multirow{2}{*}{Current algebra} & Uppal~\cite{Uppal:1994pt} & 0.3 & - \\
			
			 & Cheng~\cite{Cheng:2018hwl} & $1.28$ & - \\
			\hline
			\multirow{6}{*}{SU(3) flavor symmetry} & Sharma~\cite{Sharma:1996sc} & $0.2^{a} (1.7^{b})$ & - \\
			  & Geng~\cite{Geng:2018plk} & $1.25^{+0.38}_{-0.36}$ & -\\
			  & Geng~\cite{Geng:2018rse} & $1.30\pm0.10$ & -\\
			  & Hsiao~\cite{Zhao:2018mov} & $1.24\pm0.21$ & -\\
			  & Geng~\cite{Geng:2020zgr} & - & $0.63\pm0.34$ \\
			  & Hsiao~\cite{Hsiao:2019yur} & - & $1.14\pm0.54$\\
              & Zhong~\cite{Zhong:2022exp} & $1.36^{a}(1.27^{b})$ & -\\
              \hline
			Topological diagram method & Hsiao~\cite{Hsiao:2021nsc} & $1.42\pm0.23^{c}~ (1.47\pm0.28^{d})$ & -\\
			\hline
			Heavy quark effective theory & Singer~\cite{Singer:1996ba} & - & $0.36\pm0.02$\\
		    \hline\hline
		\end{tabular}
	\end{threeparttable}
\end{table} 

\begin{figure}[!tbp]
\centering
        \hspace{-2mm}
	\begin{subfigure}[t]{0.3\linewidth}
		\includegraphics[width=1.0\textwidth]{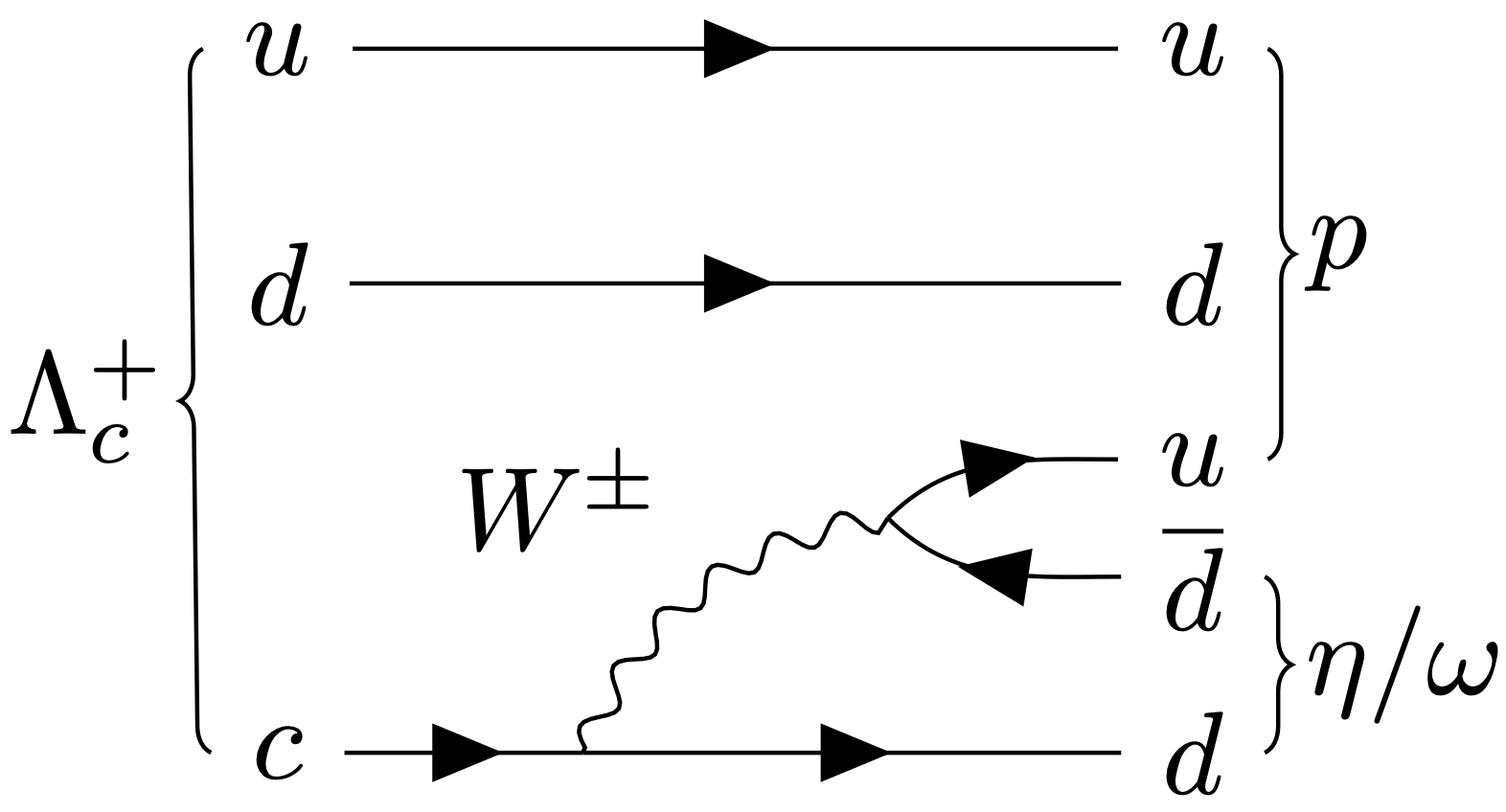}
		\captionsetup{margin={5pt, 0pt}}
		\caption{}
		\label{fig:feynman1}
		\vspace{5mm}
	\end{subfigure}
	\hspace{4mm}
	\begin{subfigure}[t]{0.3\linewidth}
		\includegraphics[width=1.0\textwidth]{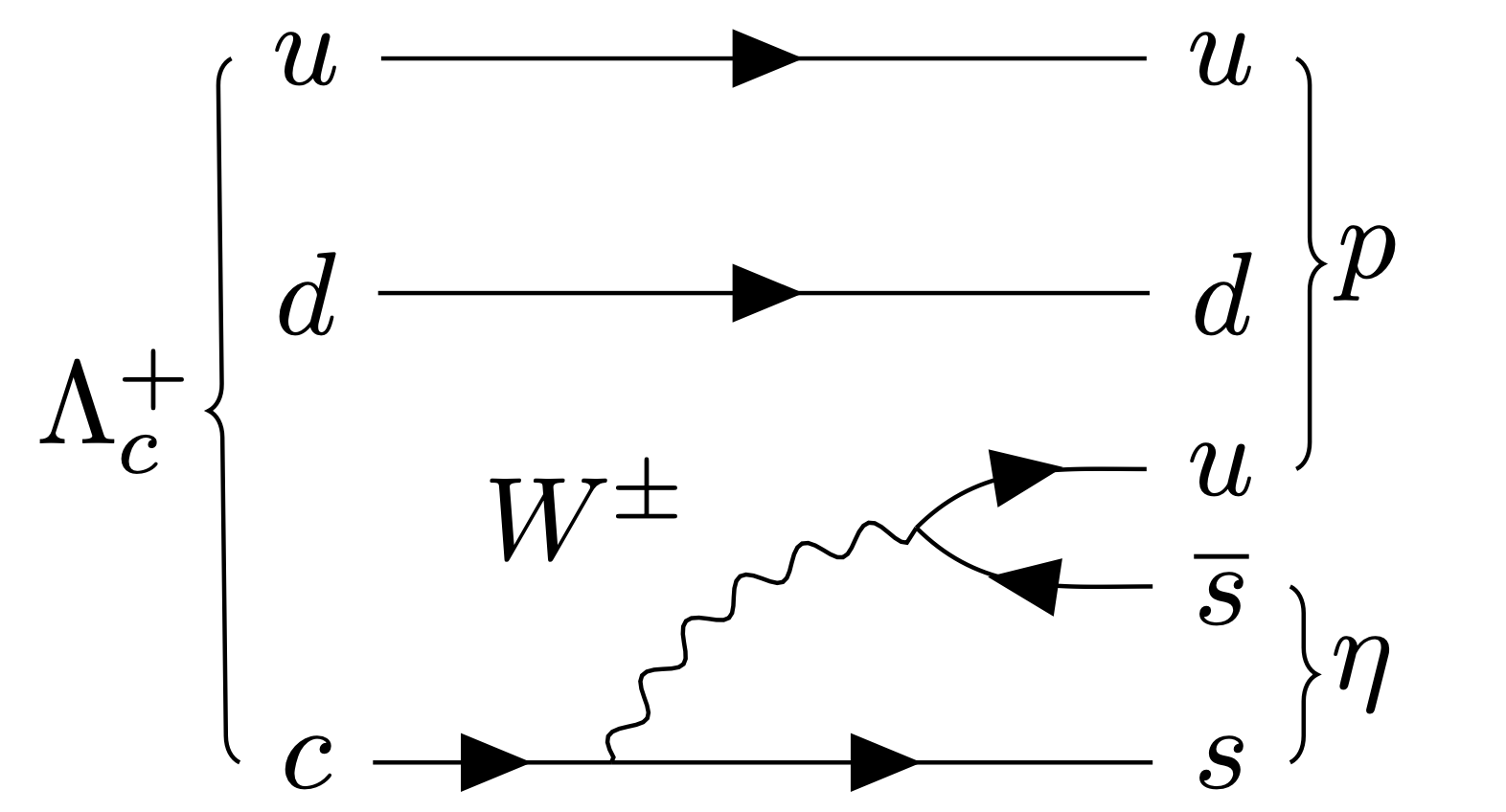}
		\captionsetup{margin={5pt, 0pt}}
		\caption{}
		\label{fig:feynman2}
		\vspace{5mm}
	\end{subfigure}
	\hspace{6mm}
	\begin{subfigure}[t]{0.3\linewidth}
		\includegraphics[width=1.0\textwidth]{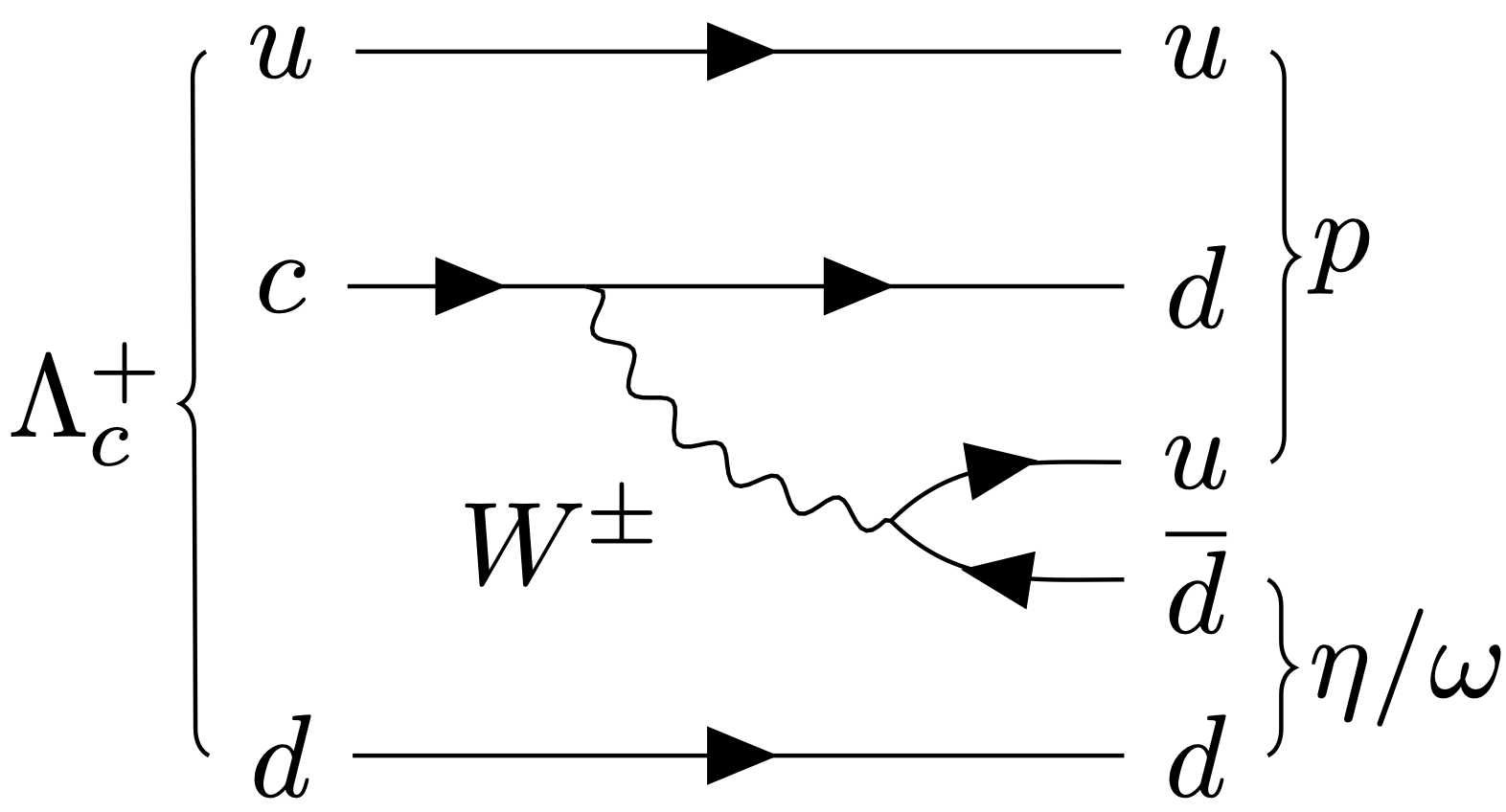}
		\captionsetup{margin={5pt, 0pt}}
		\caption{}
		\label{fig:feynman3}
		\vspace{5mm}
	\end{subfigure}
    \hspace{-15mm}
	\begin{subfigure}[t]{0.3\linewidth}
		\includegraphics[width=1.0\textwidth]{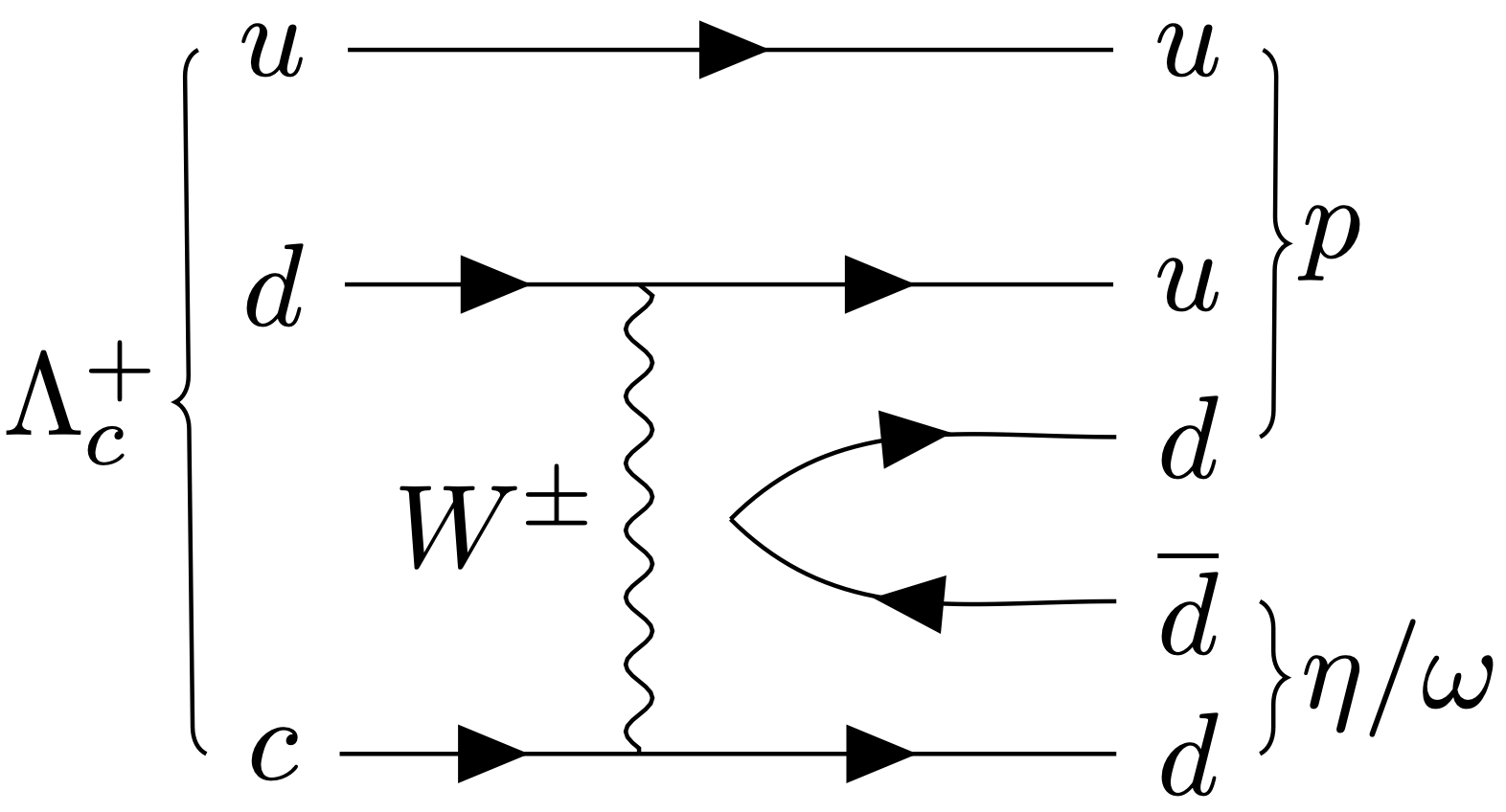}
		\captionsetup{margin={5pt, 0pt}}
		\caption{}
		\label{fig:feynman4}
	\end{subfigure}
	\hspace{5mm}
	\begin{subfigure}[t]{0.3\linewidth}
		\includegraphics[width=1.0\textwidth]{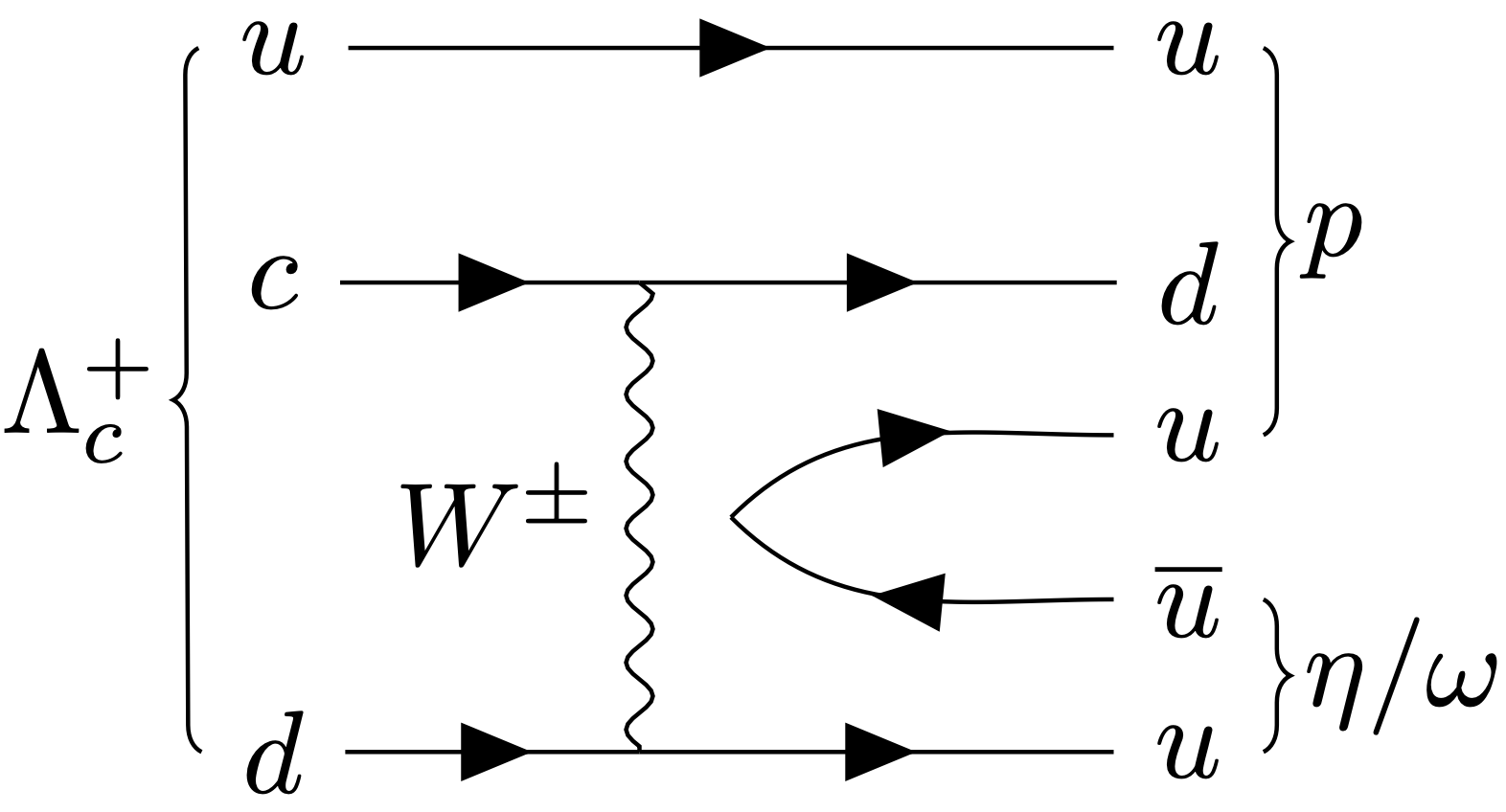}
		\captionsetup{margin={5pt, 0pt}}
		\caption{}
		\label{fig:feynman5}
	\end{subfigure}
	\hspace{5mm}
	\begin{subfigure}[t]{0.3\linewidth}
		\includegraphics[width=1.0\textwidth]{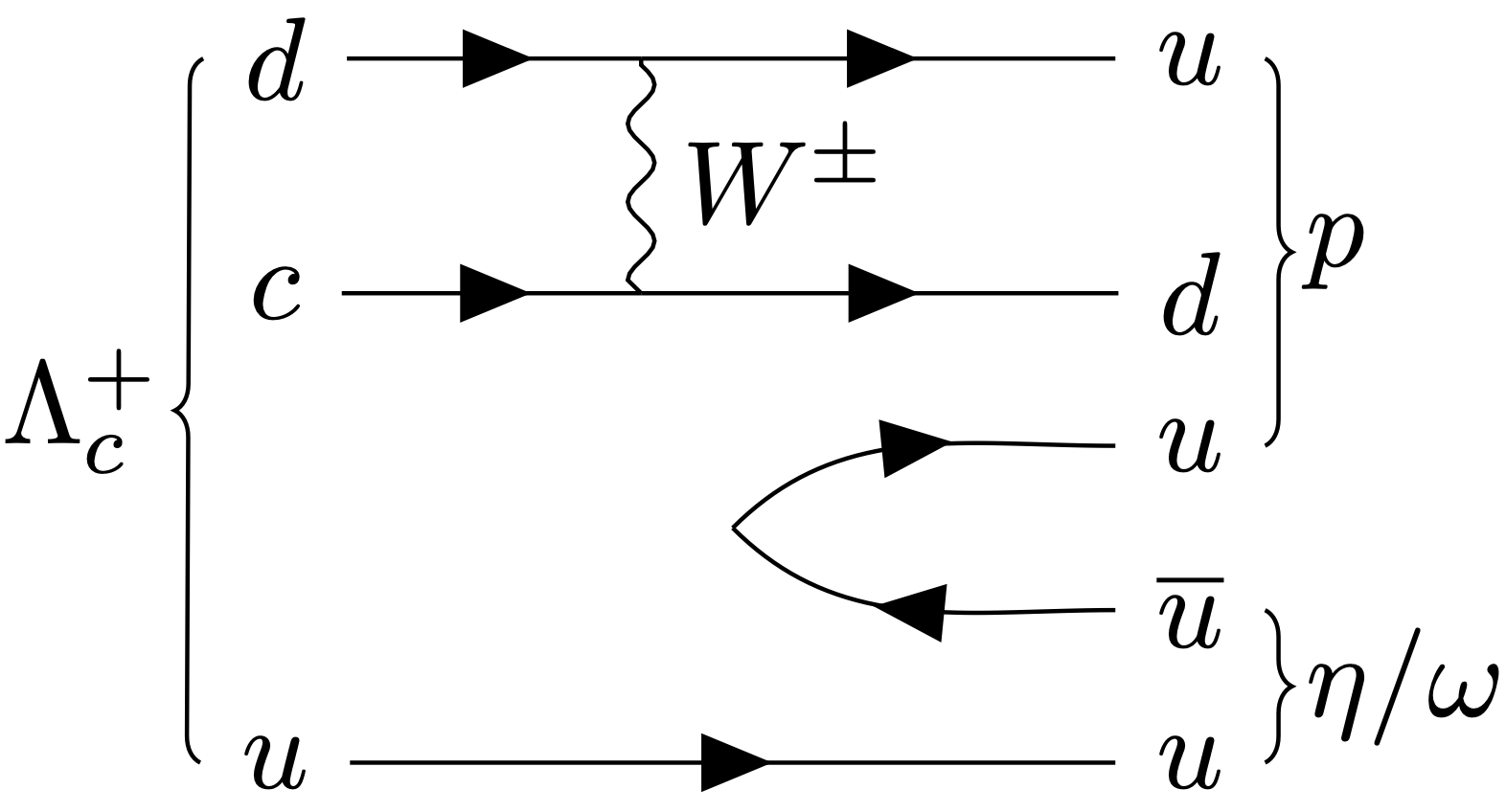}
		\captionsetup{margin={5pt, 0pt}}
		\caption{}
		\label{fig:feynman6}
	\end{subfigure}
	\caption{Feynman diagrams for $\Lambda_{c}^{+}\rightarrow p\eta/\omega$. Plots (a), (b) and (c) correspond to internal W-emission diagrams, while plots (d), (e) and (f) are W-exchange diagrams.}
	\label{fig:feynman}
\end{figure}

With respect to topological diagrams, singly Cabibbo-suppressed decays $\lcp\to p\eta$ and $\lcp\to p\omega$ occur through internal $W$-emission and $W$-exchange diagrams at tree level, as shown in \figurename~\ref{fig:feynman}. They share the same diagrams, except that $\Lambda_{c}^{+}\rightarrow p\eta$ has one additional $s$ quark involved $W$-emission amplitude in \figurename~\ref{fig:feynman2}. Diagrams of \figurename~\ref{fig:feynman1} and \figurename~\ref{fig:feynman2} are mainly factorizable, while the other diagrams in \figurename~\ref{fig:feynman} are non-factorizable.  The branching fractions of $\lcp\rightarrow p\eta$ and $p\omega$ are calculated based on various theoretical models, as listed in \tablename~\ref{tab:experiTheoPetaPomega}. In Refs.~\cite{Uppal:1994pt,Cheng:2018hwl}, the non-factorizable part of $\lcp\to p\eta$ is calculated with the pole model and the soft meson approximation, considering the parity violating amplitude. In Refs.~\cite{Sharma:1996sc, Geng:2018plk, Geng:2018rse, Zhao:2018mov, Cheng:2018hwl, Geng:2020zgr, Hsiao:2019yur, Zhong:2022exp}, global fits are carried out on the irreducible representation amplitudes based on SU(3) flavor symmetry. In Ref.~\cite{Hsiao:2021nsc}, the authors adopt the topological diagram approach, where the decay amplitudes consist of $W$-emission and $W$-exchange topologies. In Ref.~\cite{Singer:1996ba}, the branching fraction of $\lcp\to p\omega$ is predicted with the heavy quark effective theory under the factorization approximation. The theoretical results of the aforementioned phenomenological models agree with the experimental results, except for the ones in Refs.~\cite{Uppal:1994pt,Sharma:1996sc,Singer:1996ba}. Additional measurements of these two decays are necessary to improve the experimental precision and provide more stringent tests of the
different theoretical models.

In this paper, we report the measurement of the branching fractions of $\lcp\to p\eta$ and $\lcp\to p\omega$, based on the $\ee$ collision data samples collected by $\uchyph=0$BESIII ~\cite{BESIII:2009fln}
at seven energy points between 4.600 and 4.699 GeV corresponding to an integrated luminosity of $4.5\,\ifb$~\cite{BESIII:2022ulv, BESIII:2015qfd}. Charge conjugation is always implied throughout this paper, unless explicitly mentioned.

\section{BESIII Experiment and Monte Carlo Simulation}
\label{sec:detector}
\hspace{1.5em}
The BESIII detector~\cite{BESIII:2009fln} records symmetric $e^+e^-$ collisions  provided by the BEPCII storage ring~\cite{Yu:2016cof} in the center-of-mass energy range from 2.0 to 4.95~GeV, with a peak luminosity of $1 \times 10^{33}\;\text{cm}^{-2}\text{s}^{-1}$  achieved at $\sqrt{s} = 3.77\;\text{GeV}$.  The $\uchyph=0$BESIII detector has collected large data samples in this energy region~\cite{BESIII:2020nme}. The cylindrical core of the $\uchyph=0$BESIII detector covers 93\% of the full solid angle and comprises a helium-based multilayer drift chamber (MDC), a plastic scintillator time-of-flight system (TOF), and a CsI(Tl) electromagnetic calorimeter (EMC), which are all enclosed in a superconducting solenoidal magnet providing a 1.0 T magnetic field \cite{detvis}. The solenoid is supported by an octagonal flux-return yoke with resistive plate counter muon identification modules interleaved with steel.  The charged-particle momentum resolution at $1\,\gevc$ is $0.5\%$, and the $\mathrm{d}E/\mathrm{d}x$ resolution is $6\%$ for electrons from Bhabha scattering. The EMC measures photon energies with a resolution of $2.5\%$ ($5\%$) at $1\,\gev$ in the barrel (end-cap) region. The time resolution in the TOF barrel region is 68 ps, while that in the end-cap region is 110 ps. The end-cap TOF system was upgraded in 2015 using multi-gap resistive plate chamber technology, providing a time resolution of 60 ps~\cite{etof}.

Simulated data samples generated with {\sc geant4}-based~\cite{geant4} Monte Carlo (MC) software, containing the geometric description of the BESIII detector and the detector response~\cite{GDMLMethod,BesGDML}, are used to determine detection efficiencies and to estimate backgrounds. The simulation models the beam energy spread and initial state radiation (ISR) in the $e^+e^-$ annihilations with the generator {\sc kkmc}.  The inclusive MC samples include the production of $\lcp\lcm$ pairs, open charm processes, the ISR production of vector charmonium(-like) states, and the continuum QCD processes $\ee\to q\bar{q}\,(q=u,d,s)$ incorporated in {\sc kkmc}~\cite{kkmc}.  The known decay modes are modeled with {\sc evtgen}~\cite{evtgen} using branching fractions taken from the Particle Data Group (PDG)~\cite{pdg}, and the remaining unknown charmonium decays are modeled with {\sc lundcharm}~\cite{lundcharm}. Final state radiation (FSR) from charged final state particles is incorporated using {\sc photos}~\cite{photos}. Phase space (PHSP) MC samples, where the $\lcp$ decays into final states $p\eta$, $p\omega$ and $p\pip\pim\piz$ with uniform phase space distributions, are also generated. The signal MC samples are used to obtain signal shapes and estimate detection efficiencies.

\section{Event Selection}
\label{sec:selection}
\hspace{1.5em}
For $\lcp\to p\eta$, the $\eta$ is reconstructed through the two dominant modes $\eta\to\gam\gam\,(\eta_{\gam\gam})$ and $\eta\to\pip\pim\piz\,(\eta_{3\pi})$. For $\lcp\to p\omega$, the $\omega$ is reconstructed through $\omega\to\pip\pim\piz$, where neutral pions are reconstructed from two photons. Since the data sets are taken at the energy just above the $\lcp\lcm$ mass threshold, the $\lcp\lcm$ pairs are produced without any accompanying hadrons. The single-tag (ST) method is utilized in this analysis, where only one $\lcp$ is reconstructed in each event without requiring the other $\lcm$ in the recoil side. This method has a higher efficiency and allows us to acquire more $\lcp$ candidates.

Charged tracks detected in the MDC are required to be within a polar angle ($\theta$) range of $|\rm{cos\theta}|<0.93$, where $\theta$ is defined with respect to the $z$-axis, which is the symmetry axis of the MDC. Due to the short lifetime of charmed baryons ($\sim10^{-13}$~s~\cite{pdg}), charged tracks are expected to originate from the interaction point (IP). Hence, the track distance to the IP along the $z$-axis ($V_{z}$) is required to be less than 10 cm, and that perpendicular to the $z$-axis ($V_{r}$) less than 1 cm. By combining the information on the specific energy loss deposited in the MDC ($\mathrm{d}E/\mathrm{d}x$) and the time of flight measured by TOF, a likelihood ($\mathcal{L}(h)$) is calculated for each hadron hypothesis ($h=p, K, \pi$) for each charged track. Proton candidates are required to satisfy $\mathcal{L}(p)>\mathcal{L}(\pi)$, $\mathcal{L}(p)>\mathcal{L}(K)$, and $\mathcal{L}(p)>0$, while pion candidates to satisfy $\mathcal{L}(\pi)>\mathcal{L}(p)$, $\mathcal{L}(\pi)>\mathcal{L}(K)$, and $\mathcal{L}(\pi)>0$. 
The efficiency for $K/\pi$ particle identification (PID) is greater than 95\% for momentum within the region $[0.2, 1.0]\,\mathrm{GeV}/c^{2}$, where the $K/\pi$ contamination rate is less than 5\%, and the efficiency for proton PID is nearly 100\% for momentum within the region $[0.4, 1.0]\,\mathrm{GeV}/c^{2}$~\cite{BESIII:2009fln, He:2007zzj}.
A further requirement $V_{r}<0.2\,\mathrm{cm}$ is imposed on the proton candidates to avoid protons produced from beam interactions with residual gas in the beam pipe, the materials of the beam pipe or the MDC inner wall.

Photon candidates from $\piz$ and $\eta$ decays are reconstructed by electromagnetic showers produced in the EMC. The deposited energy of each shower is required to be greater than $25\,\mev$ in the barrel region ($|\cos\theta|<0.80$) and $50\,\mev$ in the end-cap region ($0.86<|\cos\theta|<0.92$). In order to suppress the showers produced by electronic noise, beam background or unrelated to the event, the difference between the EMC time and event start time~\cite{Guan:2013jua} is required to be within 700~ns. Showers are required to be separated by more than 8$^{\circ}$ from other charged tracks and more than 20$^{\circ}$ from anti-protons, to eliminate showers induced by charged tracks. The EMC shower shape variables are used to distinguish photons from anti-neutrons: the lateral moment \cite{Drescher:1984rt} should be less than 0.4 and $E_{3\times3}/E_{5\times5}$ should be greater than 0.85, where $E_{3\times3}$ ($E_{5\times5}$) is the deposited energy summed over $3\times3$ ($5\times5$) crystals around the center of the shower.

For $\piz$ ($\eta_{\gam\gam}$) candidates, the invariant mass of the photon pair is required to be within the interval $0.115\,\gevcc<M_{\gam\gam}<0.150\,\gevcc$ ($0.510\,\gevcc<M_{\gam\gam}<0.570\,\gevcc$). A one-constraint (1C) kinematic fit (KF) is performed by constraining the invariant mass of the photon pair to the nominal $\piz$ ($\eta$) mass~\cite{pdg} to improve the momentum resolution. The $\chi^{2}$ of the 1C KF is required to be less than 50 (20) for $\piz$ ($\eta_{\gam\gam}$) candidates. The momenta after the 1C KF are used in the subsequent analysis. To eliminate miscombinations that accumulate at large $|\cos\theta_{\rm decay}|$, we further require $|\cos\theta_{\rm decay}|<0.9$ for $\eta_{\gam\gam}$ candidates, where $\theta_{\rm decay}$ is the helicity angle of one photon candidate in the rest frame of the $\eta_{\gam\gam}$ candidate. For $\eta_{3\pi}$ ($\omega$) candidates, the invariant mass of the three pion combinations is required to be in the region $0.536\,\gevcc<M_{\pip\pim\piz}<0.560\,\gevcc$ ($0.750\,\gevcc<M_{\pip\pim\piz}<0.810\,\gevcc$).

For $\lcp\to p\eta_{3\pi}$ candidates, events with the invariant mass $M_{p\piz}$ within the region $(1.17, 1.20)\,\gevcc$ are rejected to suppress the intermediate $\sgmp$ contributions from $\lcp\to\sgmp\,(\to p\piz) \pip\pim$. For the $\lcp\to p\omega$ candidates, we apply a vertex fit to the proton and two charged pions, and the resultant momenta are used in further analysis.  In addition, we veto events with the invariant mass of $M_{p\piz}$, $M_{p\pim}$, and $M_{\pip\pim}$ in the regions of $(1.17, 1.20)\,\gevcc$, $(1.10, 1.12)\,\gevcc$, and $(0.47, 0.51)\,\gevcc$, respectively, to remove the contributions from $\lcp\to\sgmp\pip\pim$, $\lcp\to\Lambda\pip\piz$, and $\lcp\to p\Ks\piz$.
The amplitude for $\omega\to\pip\pim\piz$ is conventionally expanded as a polynomial around the center of the Dalitz plot, as shown in \figurename~\ref{fig:XYDalitz}, in terms of symmetrized coordinates
$X$ and $Y$~\cite{Schneider:2010hs}, which are defined as 

\begin{eqnarray} 
\label{eq:XYdef1}
X&=&\sqrt{3}(T_{+}-T_{-})/Q,\\ 
\label{eq:XYdef2} 
Y&=&3T_{0}/Q - 1,
\end{eqnarray}

\noindent where $T_{+}$, $T_{-}$, and $T_{0}$ are the kinetic energies of $\pip$, $\pim$, and $\piz$ in the rest frame of $\omega$. $Q$ is defined as $Q=M_{\omega}-(m_{\pip}+m_{\pim}+m_{\piz})$, where $m_{\pip}$, $m_{\piz}$, and $m_{\pim}$ are the nominal masses from the PDG~\cite{pdg}, and $M_{\omega}$ denotes the invariant mass of $\omega$ meson reconstructed with three pions $M_{\omega}=M_{\pip\pim\piz}$.  
The Dalitz plot with symmetrized coordinates
has advantage in deriving a variable of normalized distance $\mathcal{R}$, which is defined as 
\begin{eqnarray} 
\mathcal{R} = \sqrt{\frac{X^{2}+Y^{2}}{X_{\rm bound}^{2}+Y_{\rm bound}^{2}}}. 
\label{eq:Rdef} 
\end{eqnarray} 
Here, $(X_{\rm bound}, Y_{\rm bound})$ is the intersection of a line through the origin $(0, 0)$ and $(X, Y)$ with the boundary curve, as shown in \figurename~\ref{fig:XYDalitz}. Hence, $\mathcal{R}$ represents the scaled distance between $(0, 0)$ and $(X, Y)$. \figurename~\ref{fig:sigsbregion} shows the $\omega$ signal and the sideband regions. The signal region is defined as $0.750\,\gevcc<M_{\pip\pim\piz}<0.810\,\gevcc$, while the sideband regions as $0.620\,\gevcc<M_{\pip\pim\piz}<0.720\,\gevcc$ and $0.840\,\gevcc<M_{\pip\pim\piz} <0.890\,\gevcc$. Figures~\ref{fig:XYsig} and \ref{fig:XYsb} show the Dalitz plots of the signal MC sample and non=$\omega$ background events, respectively, where $\omega$ signals concentrate at the origin, while non-$\omega$ background events distribute uniformly.  Therefore, we require the $\mathcal{R}$ value to be less than 0.9 to further suppress the backgrounds from non-$\omega$ contribution.

\begin{figure}[thp]
\centering
\begin{subfigure}[t]{0.45\textwidth}
\includegraphics[width=1.0\textwidth]{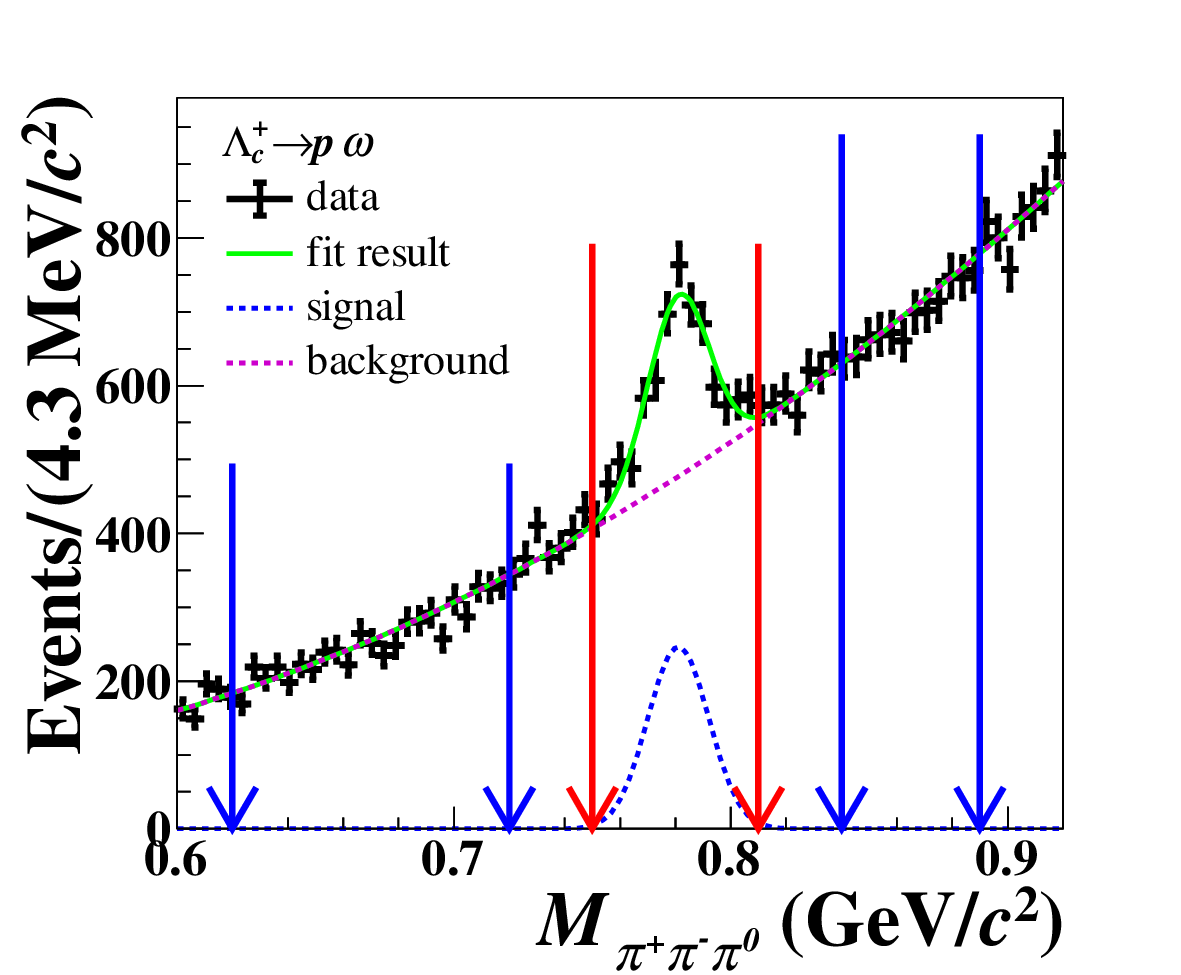}
\caption{}
\label{fig:sigsbregion}
\end{subfigure}
\begin{subfigure}[t]{0.45\textwidth}
\includegraphics[width=1.0\textwidth]{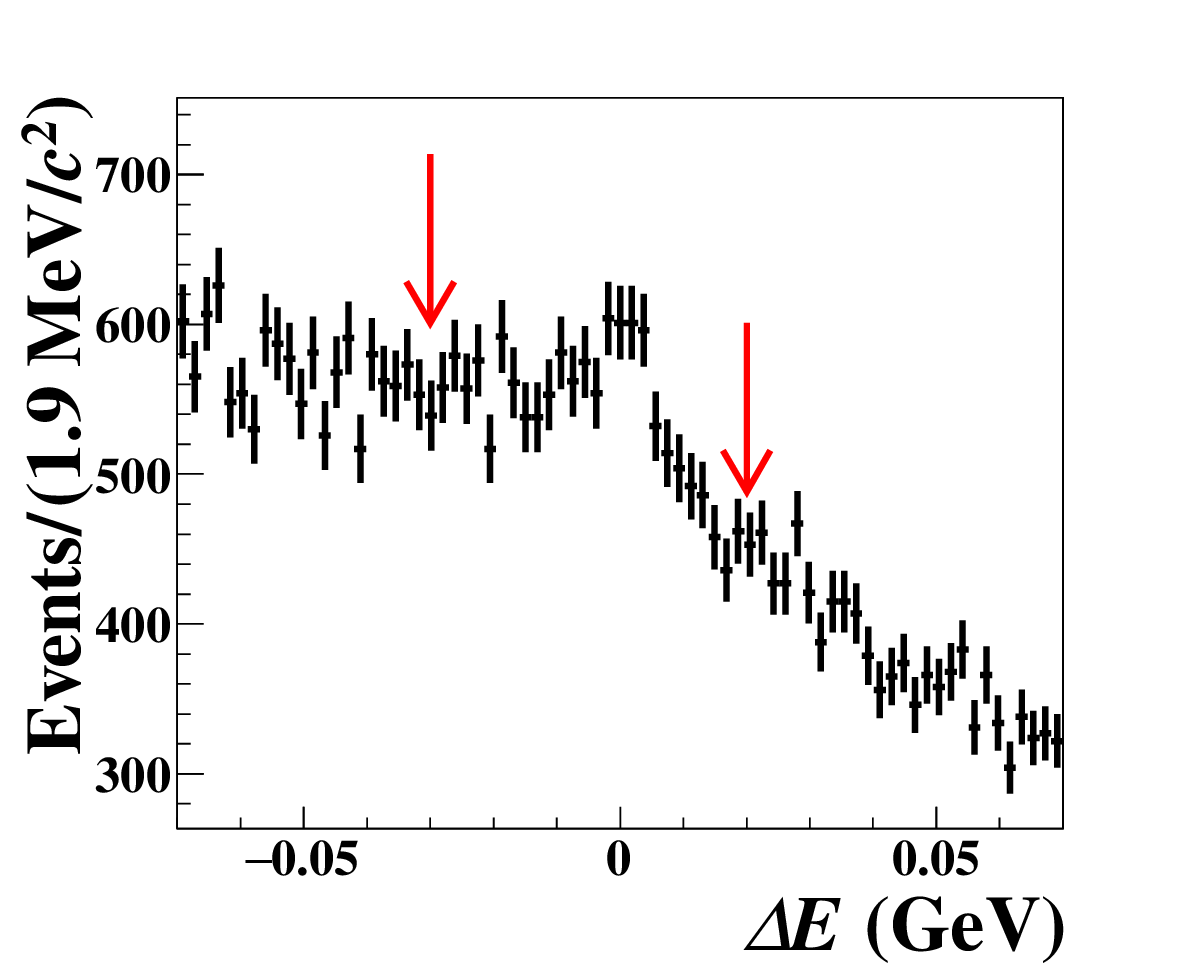}
\caption{}
\label{fig:deltae}
\end{subfigure}
\begin{subfigure}[t]{0.45\textwidth}
\includegraphics[width=1.0\textwidth]{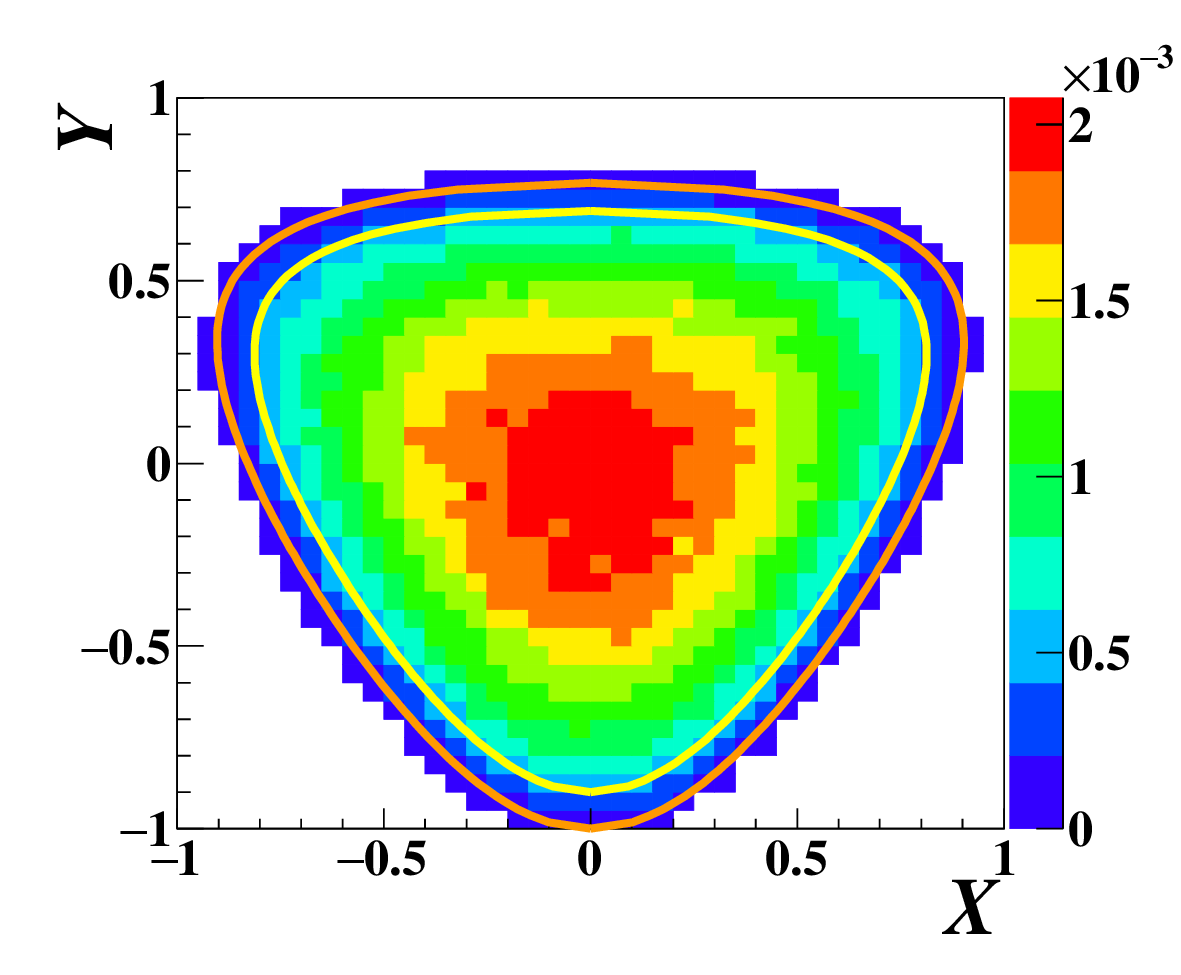}
\caption{}
\label{fig:XYsig}
\end{subfigure}
\begin{subfigure}[t]{0.45\textwidth}
\includegraphics[width=1.0\textwidth]{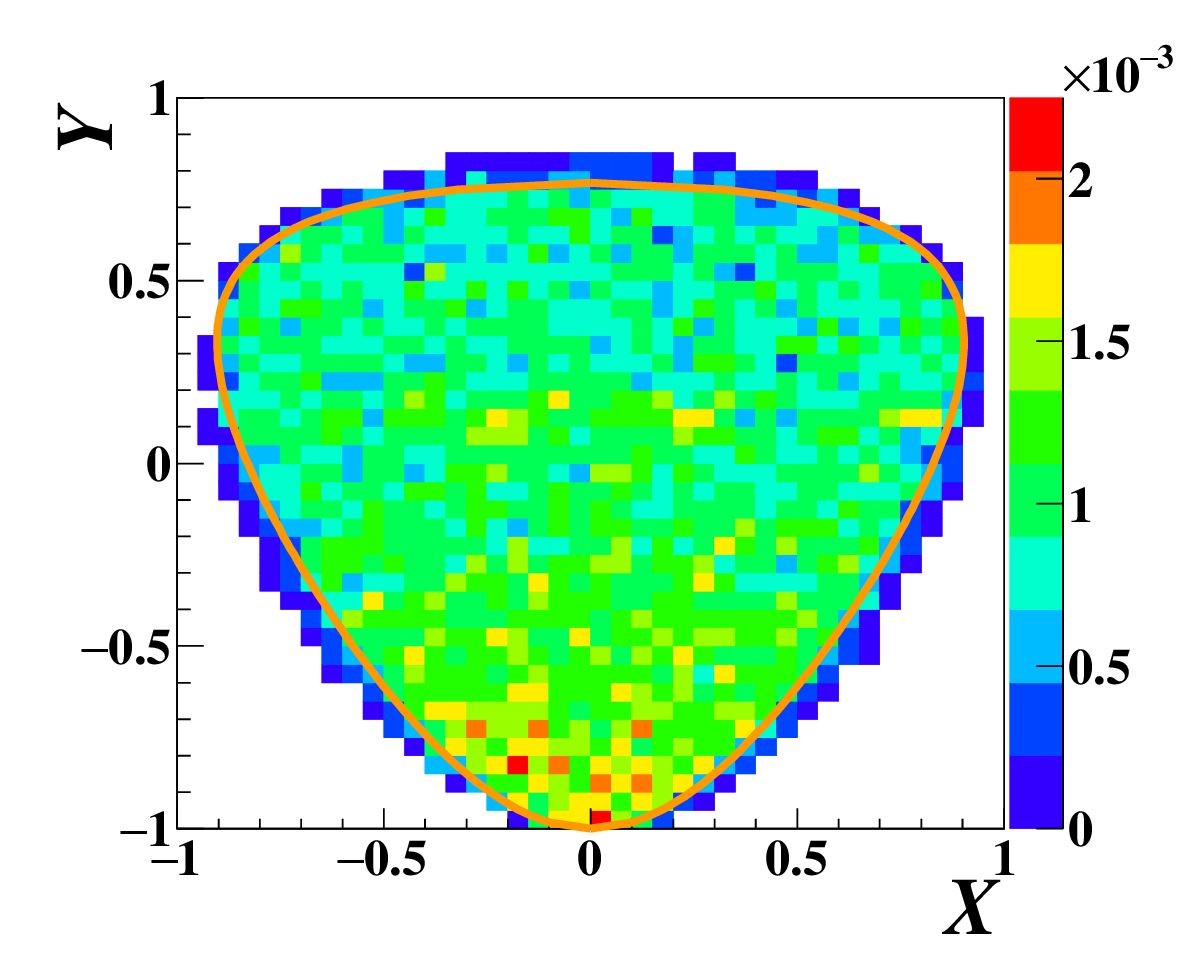}
\caption{}
\label{fig:XYsb}
\end{subfigure}
\caption{(a) Fit to the $M_{\pip\pim\piz}$ distribution, where the blue dashed line denotes the $\omega$ signal events, and the magenta dashed line are non-$\omega$ contributions. The red arrows mark the signal region, and the blue arrows mark the sideband regions. (b) $\Delta E$ distribution from data, where the red arrows mark the $\Delta E$ window requirement. (c) and (d) Dalitz plots after all requirements except the $\mathcal{R}$ requirement in the dimensions of $X$ and $Y$, as defined in Eqs.~(\ref{eq:XYdef1}--\ref{eq:XYdef2}), of the $\omega$ signal MC sample and non-$\omega$ background events from the $M_{\pip\pim\piz}$ sideband regions, respectively. The orange lines tracing the borders of Dalitz plots are the kinematic limits of $\omega\to\pip\pim\piz$. The yellow curve marks the position of the $\mathcal{R}$ requirement.}
\label{fig:XYDalitz}
\end{figure}

To further identify signal candidates, the beam constrained mass $\mbc$ and the energy difference $\dE$ variables are used, defined as
\begin{eqnarray}
	\mbc &=& \sqrt{\ebeam^{2}/\cfsq-\left|\vec{p}_{\lcp}\right|^{2}/\csq},\\
	\dE &=& E_{\lcp}-\ebeam,
	\label{eq:mbcdEdef}
\end{eqnarray} 

\noindent where $\ebeam$ is the beam energy, and $\vec{p}_{\lcp}$ and $E_{\lcp}$ are momentum and energy of the $\lcp$ candidate in the rest frame of $\ee$, respectively.  For a correctly reconstructed $\lcp$ candidate, one expects $\mbc$ to be the nominal $\lcp$ mass (2.28646 $\gevcc$~\cite{pdg}) and $\Delta E$ to be zero. The beam constrained mass of $\lcp$ candidates should satisfy $2.25~\gevcc<\mbc<\ebeam/\csq$ and $2.22~\gevcc<\mbc<\ebeam/\csq$ for $\lcp\to p\eta$ and $\lcp\to p\omega$, respectively. For the decay channel $\lcp\to p\eta$, a loose constraint on the energy difference $-0.07\,\gevcc<\dE<0.07\,\gevcc$ is used, and a two-dimensional (2D) fit to the $\mbc$ and $\Delta E$ distributions is performed to obtain the signal yield. The details will be reported in the next section. 
For the $\lcp$ reconstructed from $\lcp\to p \omega$, a tight constraint on the energy difference $-0.03\,\gev<\dE<0.02\,\gev$ is used to suppress the high $q\bar{q}$ backgrounds, as shown in \figurename~\ref{fig:deltae}. Therefore, the 2D fit to the $M_{\rm BC}$ and $\Delta E$ distributions is not applied, and the signal yield is extracted from one-dimensional (1D) fit to $M_{\rm BC}$ spectrum.

Multiple $\lcp$ candidates survive after the above selection criteria and the fractions of events with more than one candidate are listed in \tablename~\ref{tab:candFrac}. 
For $\lcp\to p\eta_{\gamma\gamma}$, the candidate with the maximum value of proton PID probability and the minimum value of $\chi^{2}$ from the 1C KF of $\eta\to\gamma\gamma$ is retained. For $\lcp\to p\eta_{3\pi}$, the candidate with the maximum value of proton PID and the minimum value of $\left|M_{\pip\pim\piz}-m_{\eta}\right|$ is kept, where $m_{\eta}$ is the nominal value of $\eta$ mass from the PDG~\cite{pdg}. For $\lcp\to p\omega$, the candidate with the minimum $|\dE|$ is selected. 

\begin{table}[!htbp]
	\centering
	\caption{The fractions of events with more than one candidate at different energy points in percentage.}
	\label{tab:candFrac}
	\begin{tabular}{c|c|c|c|c}
		\hline\hline
		$i$ & Energy Points ($\gev$) & $p\eta_{\gamma\gamma}$ & $p\eta_{3\pi}$ & $p\omega$ \\
		\hline
		1 & 4.600 & 4.86  & 12.77 & 20.06 \\
		2 & 4.612 & 4.62  & 20.51 & 16.34\\
		3 & 4.628 & 5.19  & 13.68 & 19.71\\
		4 & 4.641 & 6.05  & 15.50 & 16.14\\
		5 & 4.661 & 7.35  & 16.85 & 18.10\\
		6 & 4.682 & 7.30  & 16.52 & 18.79\\
		7 & 4.699 & 10.08 & 15.20 & 19.98\\
		\hline\hline
	\end{tabular}
\end{table}

\section{Branching Fraction Measurement}
\label{sec:measurement}
\hspace{1.5em}
The branching fractions of signal modes are determined by a simultaneous maximum likelihood fit to the data sets at seven energy points.  The branching fractions of $\lcp\to p\eta$ and $\lcp\to p\omega$ at each energy point ($i$ indicates the $i$-th energy point, from 1 to 7) is determined by 
\begin{eqnarray}
	\mathcal{B}=\frac{N_{\rm sig}^{i}}{2\cdot N_{\lcplcm}^{i}\cdot\varepsilon^{i}\cdot\mathcal{B}_{\rm inter}},
	\label{eq:bf}
\end{eqnarray}
\noindent where $N_{\rm sig}^{i}$ is the signal yield, $N_{\lcplcm}^{i}$ is the number of $\lcplcm$ pairs, $\varepsilon^{i}$ is the detection efficiency estimated according to the signal MC sample, and $\mathcal{B}_{\rm inter}$ represents the branching fractions of intermediate states, including $\eta\to\gamma\gamma$, $\eta\to\pip\pim\piz$, $\piz\to\gam\gam$, and $\omega\to\pip\pim\piz$, from the PDG~\cite{pdg}.  The numbers of $\lcplcm$ pairs for each energy point, which are listed in \tablename~\ref{tab:numLcLcPairs}, are calculated by $N_{\lcplcm}^{i}=\mathcal{L}^{i}\times\sigma^{i}$, where $\mathcal{L}_{i}$ and $\sigma_{i}$ denote the luminosity~\cite{BESIII:2022ulv} and cross section~\cite{xsec}, respectively.

\begin{table}[!htbp]\footnotesize
    \renewcommand{\arraystretch}{1.7}
	\centering
	\caption{Number of $\lcplcm$ pairs ( $N^i_{\lcplcm}$), luminosities ($\mathcal{L}^{i}$), efficiencies ($\varepsilon^{i}$), and number of signal events from simultaneous fit ($N_{\rm sig}^{i}$) and separate fit ($N^{'i}_{\rm sig}$) at different energy points.}
	\label{tab:numLcLcPairs}
	\resizebox{\linewidth}{!}{
		\begin{tabular}{c|c|c|c|c|c|c|c}
			\hline\hline
            Energy Points ($\gev$) & 4.600 & 4.612 & 4.626 & 4.640 & 4.660 & 4.680 & 4.700 \\
            \hline
            $N^i_{\lcplcm}$ & $99222\pm3671$ & $17441\pm828$ & $89302\pm3307$ & $95451\pm3535$ & $91609\pm3416$ & $278540\pm9738$ & $84342\pm3252$ \\
            $\mathcal{L}^{i}~(\mathrm{pb}^{-1})$ & $586.89\pm3.90$ & $103.65\pm0.55$ & $521.53\pm2.76$ & $551.65\pm2.81$ & $529.43\pm2.81$ & $1667.39\pm8.84$ & $535.54\pm2.84$ \\
            $\varepsilon^{i}$ for $p\eta_{\gamma\gamma}$ (\%) & $41.70\pm0.07$ & $41.58\pm0.07$ & $41.63\pm0.07$ & $41.44\pm0.07$ & $41.37\pm0.07$ & $41.34\pm0.07$ & $41.09\pm0.07$ \\
            $N_{\rm sig}^{i}$ for $p\eta_{\gamma\gamma}$ & $50\pm12$ & $9\pm5$ & $45\pm11$ & $48\pm11$ & $46\pm11$ & $140\pm19$ & $42\pm11$ \\
            $N_{\rm sig}^{'i}$ for $p\eta_{\gamma\gamma}$ & $34\pm9$ & $8\pm4$ & $36\pm10$ & $56\pm11$ & $38\pm11$ & $185\pm22$ & $36\pm12$ \\
            $\varepsilon^{i}$ for $p\eta_{3\pi}$ (\%) & $23.25\pm0.06$ & $22.49\pm0.06$ & $23.40\pm0.06$ & $22.24\pm0.06$ & $22.28\pm0.06$ & $22.34\pm0.06$ & $22.18\pm0.06$ \\
            $N_{\rm sig}^{i}$ for $p\eta_{3\pi}$ & $17\pm6$ & $3\pm2$ & $15\pm6$ & $16\pm6$ & $15\pm6$ & $47\pm10$ & $14\pm5$ \\
            $N_{\rm sig}^{'i}$ for $p\eta_{3\pi}$ & $16\pm5$ & $0\pm2$ & $4\pm4$ & $27\pm7$ & $11\pm5$ & $59\pm11$ & $8\pm5$ \\
            $\varepsilon^{i}$ for $p\omega$ (\%) & $16.82\pm0.05$ & $16.28\pm0.05$ & $15.94\pm0.05$ & $15.81\pm0.05$ & $15.54\pm0.05$ & $15.26\pm0.05$ & $14.88\pm0.05$ \\
            $N_{\rm sig}^{i}$ for $p\omega$ & $33\pm16$ & $6\pm6$ & $28\pm15$ & $30\pm15$ & $28\pm15$ & $84\pm26$ & $25\pm14$\\
            $N_{\rm sig}^{'i}$ for $p\omega$ & $50\pm13$ & $1\pm6$ & $24\pm14$ & $16\pm15$ & $36\pm17$ & $56\pm27$ & $44\pm15$\\
			\hline\hline
		\end{tabular}
    }
\end{table}

\subsection{\texorpdfstring{$\lcp\to p\eta$}{}}
\label{sec:lmdc2peta}
\hspace{1.5em}
\begin{figure}[tbh!]
    \centering
    
        \begin{subfigure}[t]{0.9\textwidth}
        	\includegraphics[width=1.0\textwidth]{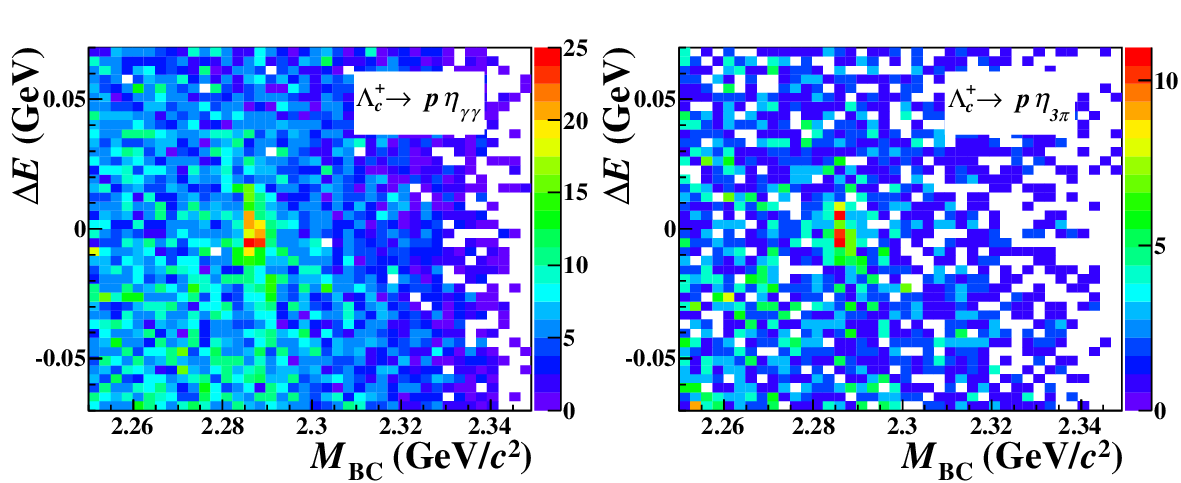} 
            \caption{}
            \label{fig:2dDataPeta}
        \end{subfigure}
        
		\begin{subfigure}[t]{0.9\textwidth}
			\includegraphics[width=1.0\textwidth]{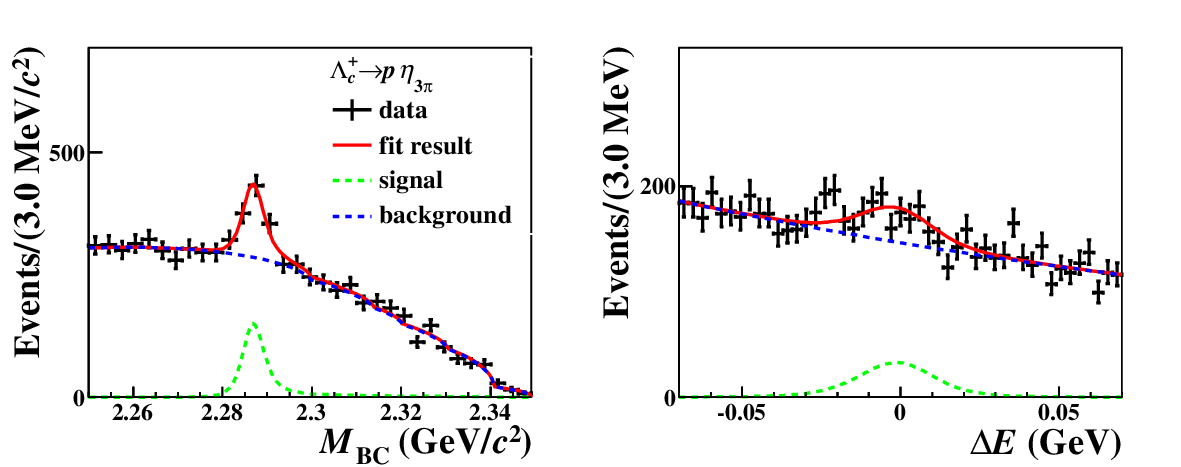}
            \caption{}
            \label{fig:1dFitDiphoton}
		\end{subfigure}
	
		\begin{subfigure}[t]{0.9\textwidth}
			\includegraphics[width=1.0\textwidth]{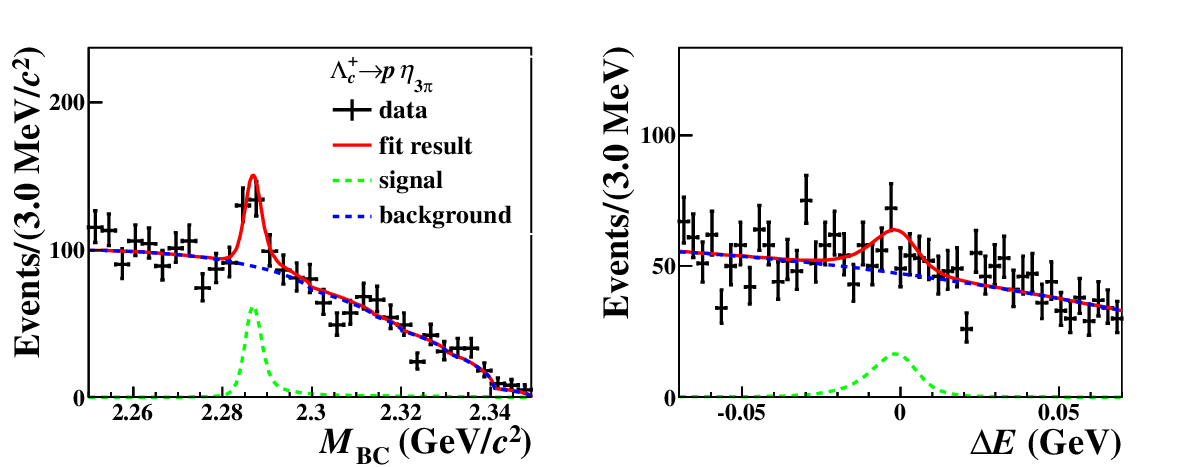}
            \caption{}
            \label{fig:1dFitTripi}
		\end{subfigure}
	
\caption{(a) 2D distributions of data summing over all the energy points.  (b, c) 1D projections of 2D simultaneous fits of $\lcp\to p\eta_{\gamma\gamma}$ and $\lcp\to p\eta_{3\pi}$ summing over the corresponding distributions at the seven energy points, respectively. Black points with error bars are data, green dashed lines are the signal shapes, blue dashed lines are background, and red solid lines are the total fitting results.}
\label{fig:fitPeta}
\end{figure}
To determine the signal yield, a 2D unbinned maximum likelihood fit is performed on the 2D distribution of $\mbc$ versus $\dE$ for each energy point.  \figurename~\ref{fig:2dDataPeta} shows 2D distributions from all data sets.  The signal shape obtained from the signal MC sample is convolved with a 2D Gaussian function to account for the resolution difference between data and MC simulation. The signal shapes among seven energy points are obtained separately.  The parameters of the 2D Gaussian are fixed in the fit to the ones obtained from the control sample of $D^{+}\to\eta\pip$ taken at $3.773~\gev$, because of the similar final states as the signal mode.  The background shape is modeled as a product of an ARGUS function \cite{ARGUS}, denoted as $f_{\rm ARGUS}$, and a second order Chebyshev polynomial function, denoted as $P_{\rm Cheb}$.  
The parameters of $f_{\rm ARGUS}$, $P_{\rm Cheb}$, and $\mathcal{B}$ are left floating in the simultaneous fit. The last one is shared by all energy points while the others are allowed to be different.
Figures~\ref{fig:1dFitDiphoton} and \ref{fig:1dFitTripi} show one-dimensional (1D) projections of the fitting results of $\lcp\to p\eta_{\gamma\gamma}$ and $\lcp\to p\eta_{3\pi}$, respectively. The branching fractions obtained via the two decay modes are consistent and are found to be $(1.55\pm0.13_{\rm stat})\times10^{-3}$ and $(1.64\pm0.21_{\rm stat})\times10^{-3}$, with a statistical significance greater than 10$\sigma$ and greater than 5$\sigma$, respectively. 
The statistical significance is evaluated by $\sqrt{-2\ln{L_{0}}/L_{\rm max}}$, where $L_{\rm max}$ is the maximum likelihood obtained from the nominal fit and $L_{0}$ is the maximum likelihood of the fit without including the signal component.
The efficiencies and signal yields from simultaneous and separate fits are given in \tablename~\ref{tab:numLcLcPairs}, where the yields from separate fits are in agreement with the yields from simultaneous fits.

\begin{figure}[tph]
    \centering
		\begin{subfigure}[t]{0.45\textwidth}
		\includegraphics[width=1.0\textwidth]{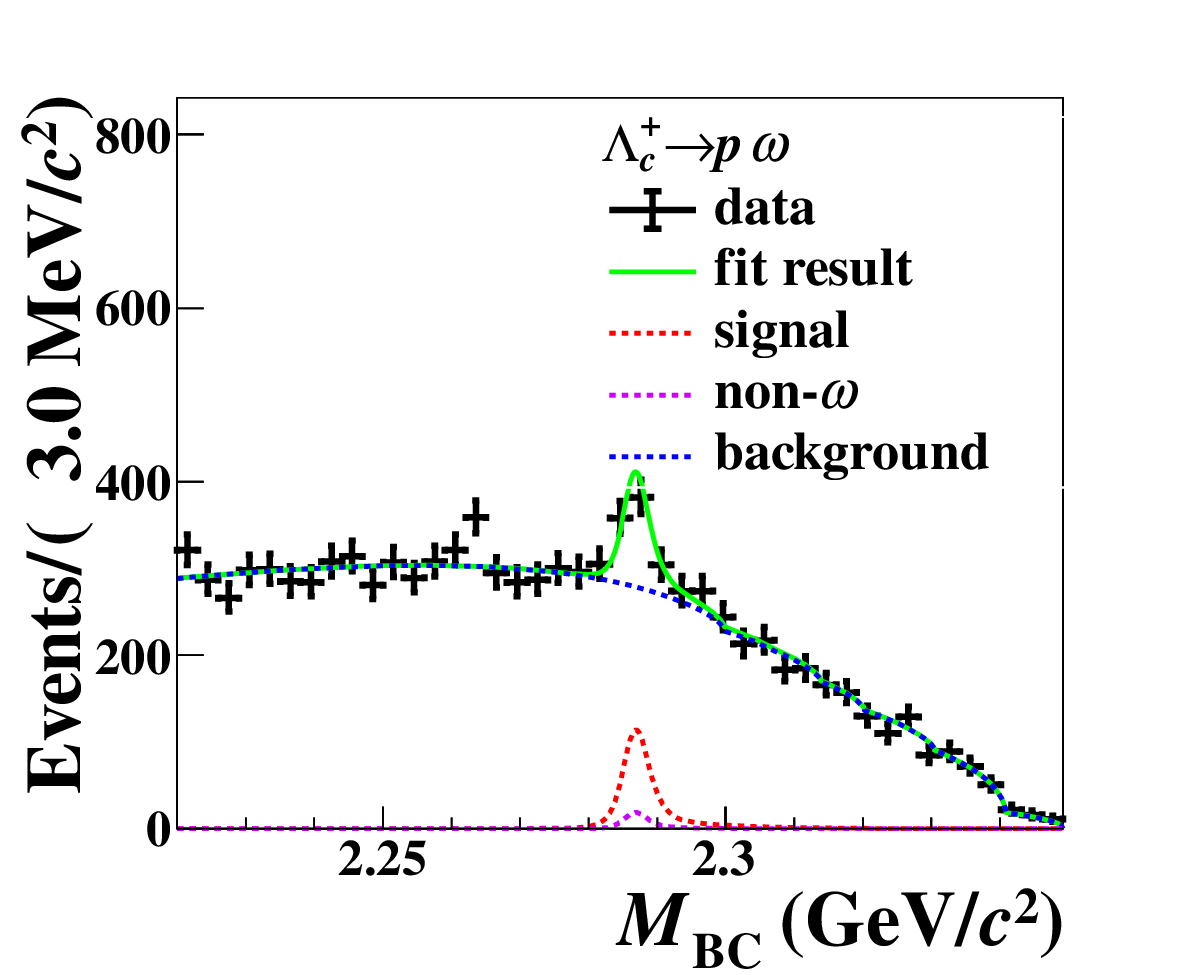}
	        \caption{}
            \label{fig:fitPomegaSig}	
        \end{subfigure}
		\begin{subfigure}[t]{0.45\textwidth}
			\includegraphics[width=1.0\textwidth]{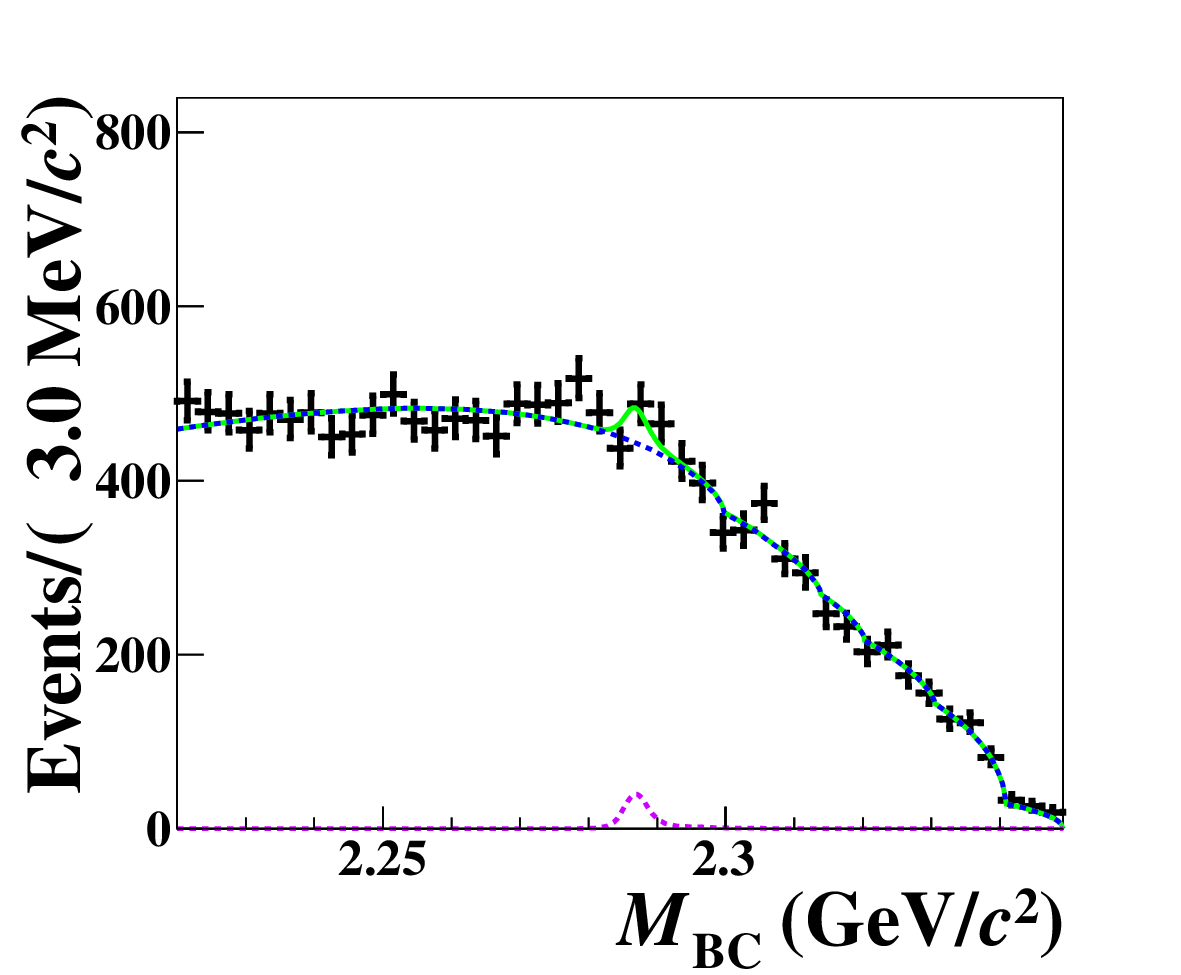}
            \caption{}
            \label{fig:fitPomegaSide}
		\end{subfigure}
\caption{Simultaneous fit to the $\mbc$ distributions for data in the $\omega$ signal (a) and sideband regions (b) as defined in \figurename~\ref{fig:sigsbregion}, summing over the corresponding distributions at seven energy points. Black dots with error bars are data, red dashed lines are signal shapes, violet dashed lines represent the non-$\omega$ peaking background contributions of $\lcp\to p \pip\pim\piz$, blue dashed lines are combinatorial background, and green solid lines are the total fitting results.}
\label{fig:fitPomega}
\end{figure}
\subsection{\texorpdfstring{$\lcp\to p\omega$}{}}
\label{sec:lmdc2pomega}
\hspace{1.5em}
A 1D unbinned maximum likelihood fit is simultaneously performed to the $\mbc$ distributions for data selected in the $\omega$ signal and sideband regions at seven energy points, as shown in Figure~\ref{fig:fitPomega}, to estimate the branching fraction of $\lcp\to p\omega$. 
The 2D fit is not used because the $q\bar{q}$ background would be too high to control if the restriction on the $\Delta E$ is released. 
The signal shape is obtained from signal MC simulations convolved with a Gaussian function. The signal shapes among seven energy points are obtained separately. The parameters of the Gaussian function are fixed according to the control sample $D^{+}\to\omega\pip$ taken at $3.773~\gev$, because of the similar final states as the signal mode.  The background is composed of two parts: the combinatorial part and the remaining non-$\omega$ peaking background. The combinatorial background is modeled by an ARGUS function, while the non-$\omega$ peaking background is modeled with the shape derived from the $\lcp\to p\pip\pim\piz$ MC sample, where its yields are constrained in the simultaneous fit to the $\lcp$ yields in the $\omega$ sideband regions in Figure~\ref{fig:fitPomegaSide}. Here, according to studies on MC simulations, other non-$\omega$ backgrounds apart from $\lcp\to p\pip\pim\piz$ are negligible after the requirements described in Sec.~\ref{sec:selection}.  
According to the fit to the $M_{\pip\pim\piz}$ spectrum, as shown in \figurename~\ref{fig:sigsbregion}, the ratio between the number of non-$\omega$ peaking background events in Figure~\ref{fig:fitPomegaSig} and the corresponding number of events in $\omega$ sideband regions in Figure~\ref{fig:fitPomegaSide} is fixed to $0.468\pm0.003$. Therefore, the fitted branching fraction of $\lcp\to p\omega$ is given as $(1.11\pm0.20_{\rm stat})\times10^{-3}$, with a statistical significance of 5.7$\sigma$. 
The efficiencies and signal yields from simultaneous and separate fits are given in \tablename~\ref{tab:numLcLcPairs}, where the yields from separate fits are in agreement with the yields from simultaneous fits.
A potential peak is seen around 2.26 GeV in the $\mbc$ distribution in Figure~\ref{fig:fitPomegaSig}, which is not observed in MC simulations and the significance of the peak is merely 2.4$\sigma$, so we treat it as statistical fluctuation in the nominal model. The systematic uncertainty caused by the peak will be discussed in Sec.~\ref{sec:sys}.

\section{Systematic Uncertainties}
\label{sec:sys}
\hspace{1.5em}
\begin{table}[!htbp]\footnotesize
    \renewcommand{\arraystretch}{1.2}
	\centering
	\caption{Systematic uncertainties from multiple sources of three signal channels in percentage. Items in bold are treated as fully correlated between $\lcp\to p\eta_{\gam\gam}$ and $\lcp\to p\eta_{3\pi}$. }
	\label{tab:sumOfSysUncer}
	\begin{tabular}{c|c|c|c}
		\hline\hline
		Sources & $\lcp\to p\eta_{\gam\gam}\,(\%)$ & $\lcp\to p\eta_{3\pi}\,(\%)$ & $\lcp\to p\omega\,(\%)$\\
		\hline
		\textbf{Proton tracking} & 0.4 & 0.4 & 0.7 \\
		\textbf{Proton PID} & 0.1 & 0.1 & 0.2 \\
		\textbf{Proton $V_r$ requirement} & 1.2 & 1.2 & 1.2\\
		Charged $\pi$ tracking & - & 0.6 & 0.6 \\
		Charged $\pi$ PID & - & 0.2 & 0.3 \\
		$\eta_{\gam\gam}\,\piz_{\gam\gam}$ reconstruction & 1.0 & 0.5 & 0.5 \\
		Shower requirements & 0.8 & 0.2 & 0.9 \\
		$M_{\pip\pim\piz}$ mass window & - & 0.2 & 1.1 \\
		$\mathcal{R}$ value & - & - & 0.5 \\
		Vetoing $\Sigma^{+}$, $K_{S}^{0}$ and $\Lambda$ & - & 1.8 & 0.7 \\
		$\Delta E$ requirement & - & - & 0.1 \\
		Non-$\eta$ and non-$\omega$ contribution & - & - & 0.7 \\
		Input $\mathcal{B}_{\rm inter}$ & 0.5 & 1.2 & 0.8 \\
		\textbf{Total number of $\lcplcm$ pairs} & 1.6 & 1.6 & 1.6 \\
		MC statistics & 0.1 & 0.1 & 0.1 \\
		MC model & 0.2 & 0.1 & 0.7 \\
		Fitting model & 0.2 & 0.5 & 5.2 \\
        \hline
        Total & \multicolumn{2}{c|}{2.4} & 6.2 \\
		\hline\hline
	\end{tabular}
\end{table}

The systematic uncertainties for $\lcp\to p\eta$ and $\lcp\to p\omega$ are summarized in \tablename~\ref{tab:sumOfSysUncer} and are described below. 

\begin{itemize} \item [\romanOne.] \emph{Proton tracking and PID.} We select a control sample of $J/\psi\to p\bar{p}\pip\pim$ to study the proton tracking and PID efficiencies as functions of proton momenta based on tag-and-based method.
The relative difference between data and MC simulations is estimated and used to reweight the nominal efficiencies in the dimensions of the signal momenta to obtain alternative efficiency.
The relative difference between the nominal efficiency and the alternative one is taken as the systematic uncertainty. This method of estimating systematic uncertainty has been used in Refs.~\cite{BESIII:2015bjk, BESIII:2022wxj}.  
The systematic uncertainties for proton tracking are evaluated to be 0.4\% for the $p\eta_{\gamma \gamma}$ and $p\eta_{3\pi}$ modes, and 0.7\% for the $p\omega$ mode; the PID uncertainties are 0.1\% for the $p\eta_{\gamma \gamma}$ and $p\eta_{3\pi}$ modes, and 0.2\% for the $p\omega$ mode.

\item[\romanTwo.] \emph{Proton $V_{r}$ requirement.}  A control sample of $\lcp\to pK^{-}\pip$ is used to study the systematic uncertainty of the $V_{r}$ requirement for proton candidates.  The difference of the estimated efficiency-corrected signal yields with and without the $V_{r}$ requirement is found to be $1.2\%$, which is taken as the systematic uncertainty.

\item [\romanThree.] \emph{Charged $\pi$ tracking and PID.}  The uncertainties due to charged $\pi$ tracking and PID are studied based on a control sample $J/\psi\to\pip\pim\piz$. We adopt the same method of estimating the systematic uncertainty as for proton tracking and PID. Accordingly, the relevant systematic uncertainties are assigned as 0.6\% (0.2\%), and 0.6\% (0.3\%) for tracking (PID) efficiencies in the $p\eta_{3\pi}$ and $p\omega$ modes, respectively.

\item [\romanFour.] \emph{$\eta_{\gamma\gamma}\,(\pi^{0}_{\gamma\gamma})$ reconstruction.} 
We study the control sample of $J/\psi\to p\bar{p}\eta_{\gamma\gamma}\,(J/\psi\to p\bar{p}\pi^{0})$~\cite{jpsi2ppbareta} to estimate the systematic uncertainty due to photon shower pairing to form $\eta_{\gamma\gamma}\,(\pi^{0})$ candidates, which is evaluated to be 1.0\% (0.5\%) with the same method as proton tracking and PID. 

\item [\romanFive.] \emph{Further $\eta$ and $\omega$ requirements.} The photon candidates from $\eta_{\gam\gam}$ and $\piz$ are chosen with additional shower requirements on the lateral moment and $E_{3\times3}/E_{5\times5}$. We further apply the helicity angle requirement, mass window requirement and the $\mathcal{R}$ value selection on $\eta_{\gam\gam}$, $\eta_{3\pi}$ and $\omega$ candidates, respectively.  The efficiencies on these additional requirements are studied by control samples of $D^{+}\to\eta_{\gamma\gamma}\pip$, $D^{+}\to\eta_{3\pi}\pip$, and $D^{0}\to\Ks\omega$.  The differences between efficiency-corrected signal yields of the control samples with and without these requirements are taken as systematic uncertainties.  The systematic uncertainty due to shower requirements and $\cos \theta_{\rm decay}$ criteria for $p\eta_{\gamma\gamma}$ is estimated to be 0.8\% in total, and that due to shower requirements and $M_{\pip\pim\piz}$ mass requirement for $p\eta_{3\pi}$ is 0.2\%. The systematic uncertainties due to shower requirements, $\mathcal{R}$ value requirement and $M_{\pip\pim\piz}$ mass window for $p\omega$ are given as 0.9\%, 0.5\% and 1.1\%, respectively.

\item[\romanSix.] \emph{Vetoing $\Sigma^{+}$, $\Ks$, and $\Lambda$.} We use control samples $\lcp\to\Sigma^{+}\pip\pim$, $\lcp\to p\Ks\piz$, and $\lcp\to\Lambda\pip\piz$ to study the data and MC difference in the $M_{p\piz}$, $M_{\pip\pim}$ and $M_{p\pim}$ spectra. The difference is described by a Gaussian function which is used to correct the signal MC sample. The relative changes on the detection efficiencies before and after correction are assigned as systematic uncertainties, which are 1.8\% and 0.7\% for
  $p\eta_{3\pi}$ and $p\omega$, respectively.

\item[\romanSeven.]\emph{$\Delta E$ requirement.}  Using the control sample of $D^{+}\to\omega\pip$, the resolution in the $\dE$ distribution in the $p\omega$ mode is studied.  The relative change of the detection efficiencies before and after correcting the MC-simulated resolution effect is assigned as the systematic uncertainty, which is 0.1\%.

\item[\romanEight.]\emph{Non-$\eta$ and non-$\omega$ contribution.} For the $p\eta_{3\pi}$ mode, a potential non-$\eta$ contribution from $\lcp\to p\pip\pim\piz$ is found to be negligible, according to a control sample in the $\eta_{3\pi}$ sideband region.  For the $p\omega$ mode, the ratio of the sizes of $\lcp\to p \pip\pim\piz$ in the $\omega$ signal and sideband regions is varied to understand the potential bias on the estimation of non-$\omega$ contributions.  An alternative method to calculate the ratio is implemented by counting the surviving numbers of events in the $\pip\pim\piz$ signal and sideband regions of the $\lcp\to p\pip\pim\piz$ MC sample. The resultant difference of the branching fractions with the new ratio from the nominal result is assigned as systematic uncertainty, which is 0.7\%.

\item[\uppercase\expandafter{\romannumeral9}] \emph{Input $\mathcal{B}_{\rm inter}$.}  The uncertainties of intermediate branching fractions are from the PDG~\cite{pdg}, and they are 0.5\%, 1.2\%, and 0.8\% for $p\eta_{\gam\gam}$, $p\eta_{3\pi}$, and $p\omega$, respectively.

\item[\uppercase\expandafter{\romannumeral10}] \emph{Number of $\lcplcm$ pairs.} The statistical uncertainties from $N_{\lcplcm}$ are 1.6\%, which include the uncertainties on the luminosity and cross section~\cite{BESIII:2022ulv,xsec} of each data set.

\item[\uppercase\expandafter{\romannumeral11}] \emph{MC statistics.} The statistical uncertainties of the estimated efficiencies from the signal MC samples are both 0.1\%.

\item[\uppercase\expandafter{\romannumeral12}] \emph{Signal model.}
In the nominal analysis, the efficiencies are calculated by using a PHSP MC sample. 
The alternative efficiencies are estimated with the weighted MC samples according to the
joint angular distributions for $p\eta$ and $p\omega$, which are written as
\begin{eqnarray}
	\mathcal{W}(\theta_{0},\theta_{1},\phi_{1})&=&1+\alpha_{0}\cdot(\cos\theta_{0})^{2}\nonumber\\
	&+&\sqrt{1-\alpha_{0}^{2}}\cdot\alpha(p\eta)\cdot\sin\Delta_{0}\cos\theta_{0}\sin\theta_{0}\sin\theta_{1}\sin\phi_{1},
\end{eqnarray}
and
\begin{eqnarray}
	\mathcal{W}(\theta_{0},\theta_{1},\phi_{1})&=&1+\alpha_{0}\cdot(\cos\theta_{0})^{2}\nonumber\\
	&+&\sqrt{1-\alpha_{0}^{2}}\cdot\frac{\beta-\alpha\gamma}{1+\gamma}\cdot\sin\Delta_{0}\cos\theta_{0}\sin\theta_{0}\sin\theta_{1}\sin\phi_{1}
\end{eqnarray}
respectively. Here, $\alpha_{0}$ and $\Delta_{0}$ denote the parameters related to the polarization of $e^+e^-\to\Lambda_{c}^{+}\bar{\Lambda}_{c}^{+}$. The parameters $\theta_{0}$ and $\theta_{1}$ denote the helicity angles of the corresponding process $e^+e^-\to\Lambda_{c}^{+}\bar{\Lambda}_{c}^{+}$ and the signal $\lcp$ decay, respectively, while $\phi_{1}$ is the angle between the initial $e^+e^-$ reaction plane and the $\lcp$ decay plane.
$\alpha(p\eta)$ denotes the asymmetry parameter of $\Lambda_{c}^{+}\to p\eta$, while $\alpha$, $\beta$ and $\gamma$ denote the asymmetry parameters of $\Lambda_{c}^{+}\to p\omega$~\cite{Wang:2016elx,Chen:2019hqi}. Due to the limited statistics of the signal yields, no practical information can be determined from data. Hence, the parameters of $\alpha(p\eta)$, $\alpha$, $\beta$ and $\gamma$ are varied within the allowed physical region, and the corresponding changes on the alternative efficiencies from the nominal efficiencies are estimated. The maximum differences are considered as systematic uncertainties, which are 0.2\%, 0.1\%, and 0.7\% for $p\eta_{\gam\gam}$, $p\eta_{3\pi}$ and $p\omega$ modes, respectively.

\item[\uppercase\expandafter{\romannumeral13}] \emph{Fitting model.}
The systematic uncertainty from the fitting model results from the signal and background shapes. To estimate the potential effects, we vary the smearing Gaussian parameters within the uncertainties from the control samples, and vary the parameter $E_{\rm beam}$ of the ARGUS function by $\pm0.2\,\mev$. Six thousand pseudo data sets are generated randomly, where for each pseudo data set these parameters are varied randomly. The pull distribution of the fitted branching fractions in the pseudo data sets indicates a relative shift of 0.2\%, 0.5\%, and 2.2\% on the results of $p\eta_{\gam\gam}$, $p\eta_{3\pi}$ and $p\omega$ modes, respectively, which are assigned as systematic uncertainties. 
For $\lcp\to p\omega$, the peak around 2.26 GeV in the $M_{\rm BC}$ distribution in Figure~\ref{fig:fitPomegaSig} is insignificant in data and not reproduced in MC simulations. However, the peak could be caused by some unknown process in truth. To estimate the potential systematic uncertainty in the nominal fit, an alternative fitting model is considered, which assumes the peak around 2.26 GeV with a Gaussian function. Two thousands toy data samples are generated based on the alternative fitting model. We fit the toy data samples with the nominal model, and the deviation of the pull distribution is taken as systematic uncertainty, which is 4.7\%.

\end{itemize}

We combine the measured branching fractions of $\lcp\to p\eta_{\gamma\gamma}$ and $\lcp\to p\eta_{3\pi}$ considering correlations of systematic uncertainties utilizing BLUE (Best Linear Unbiased Estimate)~\cite{Blue}. We assume the uncertainties from proton tracking, PID, $V_{r}$ requirement and total number of $\lcplcm$ pairs are 100\% correlated, and the other sources of systematic uncertainties are uncorrelated. The branching fraction of $\lcp\to p\eta$ is estimated to be $(1.57\pm0.11_{\rm stat}\pm0.04_{\rm syst})\times10^{-3}$.  For $\lcp\to p\omega$, we add the systematic uncertainties in quadrature, and the branching fraction is calculated to be $(1.11\pm0.20_{\rm{stat}}\pm0.07_{\rm{syst}})\times10^{-3}$.

\section{Summary}
\label{sec:summary}
\hspace{1.5em} Based on $\ee$ collision samples with an integrated luminosity of 4.5 $\mbox{fb$^{-1}$}$ collected with the BESIII detector at seven energy points between 4.600 and 4.699 GeV, the branching fractions of $\lcp\to p\eta$ and $\lcp\to p\omega$ are measured using the single tag method, and they are found to be $(1.57\pm0.11_{\rm {stat}}\pm0.04_{\rm{syst}})\times10^{-3}$ and $(1.11\pm0.20_{\rm{stat}}\pm0.07_{\rm{syst}})\times10^{-3}$, with a statistical significance greater than 10$\sigma$ and 5.7$\sigma$, respectively.  Our results are consistent with previous measurements, as given in \tablename~\ref{tab:experiTheoPetaPomega}. The result of $\lcp\to p\eta$ is the most precise single measurement to date. 
The results allow more stringent tests of various phenomenological models, as listed in \tablename~\ref{tab:experiTheoPetaPomega}, where early calculations in Refs.\cite{Uppal:1994pt,Sharma:1996sc,Singer:1996ba} are confirmed as inconsistent with experimental measurements.


~\\\\
\noindent \textbf{Acknowledgements}\\\\
The BESIII Collaboration thanks the staff of BEPCII and the IHEP computing center for their strong support. This work is supported in part by National Key R\&D Program of China under Contracts Nos. 2020YFA0406400, 2020YFA0406300; National Natural Science Foundation of China (NSFC) under Contracts Nos. 11635010, 11735014, 11835012, 11935015, 11935016, 11935018, 11961141012, 12022510, 12025502, 12035009, 12035013, 12061131003, 12192260, 12192261, 12192262, 12192263, 12192264, 12192265, 12221005, 12005311; the Chinese Academy of Sciences (CAS) Large-Scale Scientific Facility Program; the CAS Center for Excellence in Particle Physics (CCEPP); Joint Large-Scale Scientific Facility Funds of the NSFC and CAS under Contract No. U1832207; CAS Key Research Program of Frontier Sciences under Contracts Nos. QYZDJ-SSW-SLH003, QYZDJ-SSW-SLH040; 100 Talents Program of CAS; Fundamental Research Funds for the Central Universities, Lanzhou University, University of Chinese Academy of Sciences; The Institute of Nuclear and Particle Physics (INPAC) and Shanghai Key Laboratory for Particle Physics and Cosmology; ERC under Contract No. 758462; European Union's Horizon 2020 research and innovation programme under Marie Sklodowska-Curie grant agreement under Contract No. 894790; German Research Foundation DFG under Contracts Nos. 443159800, 455635585, Collaborative Research Center CRC 1044, FOR5327, GRK 2149; Istituto Nazionale di Fisica Nucleare, Italy; Ministry of Development of Turkey under Contract No. DPT2006K-120470; National Research Foundation of Korea under Contract No. NRF-2022R1A2C1092335; National Science and Technology fund of Mongolia; National Science Research and Innovation Fund (NSRF) via the Program Management Unit for Human Resources \& Institutional Development, Research and Innovation of Thailand under Contract No. B16F640076; Polish National Science Centre under Contract No. 2019/35/O/ST2/02907; The Royal Society, UK under Contract No. DH160214; The Swedish Research Council; U. S. Department of Energy under Contract No. DE-FG02-05ER41374.

\newpage
\section*{The BESIII collaboration}
\addcontentsline{toc}{section}{The BESIII collaboration}
\begin{center}

M.~Ablikim$^{1}$, M.~N.~Achasov$^{13,b}$, P.~Adlarson$^{75}$, R.~Aliberti$^{36}$, A.~Amoroso$^{74A,74C}$, M.~R.~An$^{40}$, Q.~An$^{71,58}$, Y.~Bai$^{57}$, O.~Bakina$^{37}$, I.~Balossino$^{30A}$
% ...
\\
\vspace{0.2cm}
(BESIII Collaboration)\\
\vspace{0.2cm} {\it
$^{1}$ Institute of High Energy Physics, Beijing 100049, People's Republic of China\\
% ...
$^{13}$ G.I. Budker Institute of Nuclear Physics SB RAS (BINP), Novosibirsk 630090, Russia\\
% ...
$^{30}$ INFN Sezione di Ferrara, (A)INFN Sezione di Ferrara, I-44122, Ferrara, Italy; (B)University of Ferrara,  I-44122, Ferrara, Italy\\
% ...
$^{36}$ Johannes Gutenberg University of Mainz, Johann-Joachim-Becher-Weg 45, D-55099 Mainz, Germany\\
$^{37}$ Joint Institute for Nuclear Research, 141980 Dubna, Moscow region, Russia\\
% ...
$^{40}$ Liaoning Normal University, Dalian 116029, People's Republic of China\\
% ...
$^{57}$ Southeast University, Nanjing 211100, People's Republic of China\\
$^{58}$ State Key Laboratory of Particle Detection and Electronics, Beijing 100049, Hefei 230026, People's Republic of China\\
% ...
$^{71}$ University of Science and Technology of China, Hefei 230026, People's Republic of China\\
% ...
$^{74}$ University of Turin and INFN, (A)University of Turin, I-10125, Turin, Italy; (B)University of Eastern Piedmont, I-15121, Alessandria, Italy; (C)INFN, I-10125, Turin, Italy\\
$^{75}$ Uppsala University, Box 516, SE-75120 Uppsala, Sweden\\
% ...
$^{b}$ Also at the Novosibirsk State University, Novosibirsk, 630090, Russia\\
% ...

}
   
\end{center}

\end{document}